\documentclass[usenatbib]{article}
\pdfoutput=1
\usepackage{natbib}
\usepackage{graphicx}
\usepackage{epsf,amsfonts,amsmath,amssymb}
\usepackage{mathrsfs}
\usepackage[usenames,dvipsnames,svgnames,table]{xcolor}
\usepackage{color}
\usepackage{array}
\usepackage{epstopdf}
\usepackage{multirow}
\usepackage{rotating}

\newcommand       \apj          {ApJ}
\newcommand       \apjl         {ApJL}
\newcommand       \aap          {A\&A}

\newcommand       \prc          {PRC}
\newcommand       \nat          {Nature}
\newcommand       \mnras        {MNRAS}

\newcommand       \aj      {AJ}
\newcommand       \prd      {Phys.~Rev.~D.~}

\newcommand       \araa      {ARA\&A}

\newcommand       \pasj   {PASJ}
\newcommand      \apjs {ApJ Supplements}

\newcommand \physrep{Phys.~Rep.}
\newcommand \aplett{Astrophys. Lett.}

\begin{document}

\markboth{K. Hotokezaka, P. Beniamini \& T. Piran}
{Neutron star mergers, $r$-process \& sGRB}

\title{Neutron Star Mergers as sites of $r$-process Nucleosynthesis and Short Gamma-Ray Bursts}

\author{Kenta Hotokezaka$^1$, Paz Beniamini$^2$, and Tsvi Piran$^3$\\ \\
$^1$Department of Astrophysical Sciences, Peyton Hall,\\ Princeton University, Princeton, NJ 08544, USA\\
$^2$Department of Physics, The George Washington University,\\ Washington, DC 20052, USA\\
$^3$Racah Institute of Physics, The Hebrew University of Jerusalem,\\ Jerusalem 91904, Israel
}

\maketitle

\begin{abstract}
Neutron star mergers have been  long considered as promising sites of heavy $r$-process nucleosynthesis. 
We overview observational evidence supporting this  scenario including: the total amount of $r$-process elements in the Galaxy, extreme metal poor stars, geological radioactive elemental abundances, dwarf galaxies, and short gamma-ray bursts (sGRBs). Recently, the advanced LIGO and Virgo observatories discovered a gravitational-wave signal of a neutron star merger, GW170817, as well as accompanying multi-wavelength electromagnetic (EM) counterparts. The ultra-violet, optical, and near infrared observations point to $r$-process elements that have been synthesized in the merger ejecta. The rate and ejected mass inferred from GW170817 and the EM counterparts are consistent with other observations. We find however that,  within simple one zone chemical evolution models (based on merger rates with reasonable delay time distributions as expected from evolutionary models, or from observations of sGRBs), it is difficult to reconcile the current observations of the europium abundance history of Galactic stars for [Fe/H] $\gtrsim -1$. This implies that to account for the role of mergers in the Galactic chemical evolution, we need a Galactic model with multiple populations  that have different spatial distributions and/or varying formation rates. 
\end{abstract}

\section{Introduction}
\label{sec:Introduction}
Rapid neutron capture process (``$r$-process'') has been known as a basic formation process
for the heaviest elements in the cosmos since the seminal work of B$^2$FH \citep{burbidge57} 
and \cite{cameron57}. The question in which astrophysical environment it 
actually occurs, however, has puzzled astrophysics ever since. For a long time core-collapse supernovae (cc-SNe)  have been considered as the favored production site (\citealt{takahashi94,woosley94,hoffman97,freiburghaus99a,qian1996ApJ,Thompson2001,Fryer2006,farouqi10,Arcones2011,wanajo2013ApJ,Goriely2016} and see also \citealt{qian2007PhR,arnould2007PhR,arcones13a} for   reviews). 
As an alternative site, \cite{lattimer74,lattimer1976ApJ} suggested the decompression of cold  neutron-rich matter ejected during the tidal disruption of a neutron star
by a stellar mass black hole, a process that occurs in a similar way
during the merger of two neutron stars 
(\citealt{symbalisty1982ApL}, see also \citealt{Thielemann2017} for a recent review). 

\cite{eichler89} placed compact binary mergers in a broader astrophysical context suggesting that 
Gamma-ray Bursts (GRBs), that at the time were believed to be of galactic origin, are extragalactic. Furthermore, they suggested that  in addition 
to bursts of gravitational waves and neutrinos and the production of (heavy) $r$-process material these mergers  are  the engines of  GRBs.   Today it is believed that indeed the subclass of short duration bursts (sGRBs) arise from mergers \citep{Nakar07,Berger14}.  

\cite{eichler89}  estimated the rate of mergers and  suggested that these events could be major sources of $r$-process material.  In the first detailed calculation of mass ejection from a merger, \cite{rosswog99} found that $\sim 10^{-2}M_{\odot}$ become unbound during a merger 
event due to gravitational torques and hydrodynamic interaction (this is 
referred to as ``dynamical ejecta'' in the following).  A follow-up
work  by \cite{freiburghaus99b}   performed the first nuclear network calculations based on
hydrodynamics simulations showing that the resulting abundance patterns 
agree well with the observed solar system abundances for $A > 130$ and, folding the ejecta masses with the estimated merger rates,  this indicated  that, indeed,  neutron star mergers could represent a major source of  cosmic $r$-process elements.

At the same time, \cite{LP98} suggested that  the radioactive decay of  the neutron-rich
nuclei in these ``dynamical ejecta'' would produce a macronova (also 
referred to as kilonova): a short lived optical - near infrared (nIR) weak  supernova-like signal. The heating rate of radioactive decay of r-process elements was  addressed by \cite{metzger10a}.
Studies of kilonova/macronova evolution revealed its prospects for detectability 
(\citealt{LP98,kulkarni05,metzger10a,roberts11,goriely11a,
piran13a,barnes13a,tanaka13a,grossman13a}, see also \citealt{Tanaka2016} and \citealt{Metzger2017} for reviews).
 The large opacities of lanthanides produced via $r$-process nucleosynthesis lead to the conclusion that the main emission will be in the  nIR  band peaking  a week
later than the merger \citep{kasen13a,barnes13a,tanaka13a}. Observation of such a nIR signal following a sGRB is by itself 
an indication that a significant amount of heavy $r$-process material was produced in this event. 

On June 3rd 2013, the {\it Swift} satellite detected a relatively nearby  
sGRB \citep{Melandri13} 
at a redshift of 0.356. 
At 9 days  after the burst
($\approx 6.6$ days  in the local rest frame), the  {\it Hubble Space Telescope} 
detected a nIR point source with an apparent magnitude of 
$H_{\rm 160,AB} =25.73 \pm 0.2$ ({$M_{\rm J,AB}\approx-15.35$}; \citealt{Tanvir13,Berger13}), corresponding 
to  an intrinsic luminosity of $\approx 10^{41}$\,erg/s. The upper limit on the  
r-band emission at the same time, $R_{\rm 606,AB} > 28.5$, suggests that the 
regular afterglow has decayed by this time. The  nIR excess at 9 days 
after the burst  was interpreted by both groups as
tentative evidence for a Li-Paczynski kilonova/macronova \citep{Tanvir13,Berger13}.

Later on \cite{Yang+15} found an indication of a kilonova/macronova-like bump in the archival data of the afterglow of GRB 060614 and \cite{Jin+16} found another one in the afterglow of the sGRB 050709, the first sGRB from which an optical afterglow was observed \citep{Fox2005}. \cite{Jin+16} explored other sGRBs and concluded that kilonova/macronova candidates were observed in three out of three (or at most six) possible sGRBs that were sufficiently nearby and for which data were available. 
In addition, the sGRB 080503 was accompanied by an optical transient on a time scale of a few days \citep{perley09}. Although the redshift of this burst is unknown, the optical 
bump can be consistent with a macronova model assuming $z=0.25$ \citep{kasen2015MNRAS}.   More recently, the nearby sGRB 160821B was found  at  $z=0.16$.
\cite{kasliwal2017ApJ} reported a K$_{\rm s}$-band detection at $4.3$ days after the burst, which might  arise from a kilonova/macronova. However,
a detailed modeling of the afterglow using the multi-band data is  necessary to interpret the origin of the emission.

During the past decade much effort has been invested in understanding
the relevance and the implications of mergers to  $r$-process (optical-nIR emissions;~\citealt{LP98,metzger10b,roberts11,barnes13a,grossman13a,kasen13a,piran13a,tanaka13a,kasen2015MNRAS,barnes2016ApJ,kawaguchi2016ApJ,Hotokezaka2017,Wollaeger2017,Tanaka2017arXiv},
merger and mass ejection simulations;~\citealt{rosswog00,Shibata2000,ruffert01,rosswog05a,oechslin07a,metzger08,Dessart2009,Rezzolla2011,East2012,bauswein2013ApJ,hotokezaka13b,rosswog13a,rosswog13b,metzger2014MNRAS,fernandez2013MNRAS,
2014,foucart2014PRD,fernandez2015MNRAS,just2015MNRAS,Kawaguchi2015,Kiuchi2015,Kyutoku2015,sekiguchi2015PRD,radice2016MNRAS,Ruiz2016,Bovard2017,Ciolfi2017,Dietrich2017,Foucart2017,fujibayashi2017,Kyutoku2017,Shibata2017,Siegel2017}, and 
 nucleosynthesis studies;~
\citealt{surman08,lee09, goriely11a,korobkin12a,caballero12,wanajo2012ApJ,malkus12,surman13a,wanajo2014ApJ,eichler2015ApJ,lippuner2015ApJ,wu2016MNRAS,lippuner2017}). 
The picture
that emerges from this wealth of studies is the following: a) the dynamical ejecta 
with their extreme neutron-richness provide excellent conditions for heavy ($A\gtrsim 100$) 
$r$-process and the resulting abundance pattern is largely independent
of the specifics of the merging binary system \citep{goriely11a,korobkin12a,wanajo2014ApJ,eichler2015ApJ,Bovard2017} and b) mergers
provide additional nucleosynthesis channels such as neutrino-driven winds 
and the final dissolution of the accretion disk which is likely complement of the 
nucleosynthesis from the dynamical ejecta \citep{Dessart2009,fernandez2013MNRAS,Perego2014,fernandez2015MNRAS,just2015MNRAS,wu2016MNRAS,lippuner2017,fujibayashi2017,Shibata2017,Siegel2017}.
While details of the resulting isotopic abundances vary somewhat among different models, 
the overall picture of an expected robust $r$-process pattern that agrees with the observed abundance peaks arises in most calculations. 

On 17th of August in 2017,  a gravitational-wave signal from 
a binary neutron star merger was detected by the LIGO/Virgo detectors. 
Two seconds later, a weak short-duration $\gamma$-ray burst was found  in 
the gravitational-wave localization area by the {\it Fermi} and {\it INTEGRAL} satellites \citep{gammaGW170817,Goldstein2017,Savchenko2017}.
Although this burst, GRB 170817A, is extremely weak compared to short GRBs (the isotropic 
equivalent $\gamma$-ray energy is even smaller  than the weakest short GRB by more than two orders of
magnitude), 
the observed spectra provide the first  evidence that a relativistic jet is produced by the central remnant  after  merger \citep{Gottlieb2017, Bromberg2017}.
An optical-nIR counterpart, AT2017gfo, was discovered at $\sim 11$\,hr after the merger \citep{Swope, Soares-Santos2017, Valenti2017, Arcavi2017Natur, Tanvir2017, Lipunov2017}.
 A number of follow-up observations were conducted and the photometric light curves including the ultra-violet bands and
 spectral evolution were taken \citep{Andreoni2017,Arcavi2017Natur,Chornock2017,Swope,Cowperthwaite2017,Drout2017,Evans2017,Kasliwal2017,Kilpatrick2017,Lipunov2017,McCully2017,Nicholl2017,Pian2017,Shappee2017,Smartt2017,Tanvir2017,Troja2017, Utsumi2017,Valenti2017}.
The observed features are broadly consistent with a kilonova/macronova model,
indicating that $r$-process elements with a mass of {\bf  $\sim 0.05M_{\odot}$ } have been 
synthesized in this event.   Although no specific elements are identified
in the signal, this supports 
that $r$-process elements are synthesized in merger ejecta and mergers predominately produce
the cosmic $r$-process elements. 
 At later times, X-ray \citep{Troja2017,Evans2017,Haggard2017,Margutti2017}  and radio counterparts 
 \citep{Hallinan2017,Alexander2017,mooley2017} were
also found at the same location of the optical-nIR counterpart. These non-thermal emissions arise most likely from the forward shock between the ejected outflow and interstellar medium (ISM). It is likely that  the outflow producing this emission has been energized by a choked jet that has formed a cocoon. 
The radio and X-ray emissions around 100 days require a mildly relativistic outflow and this  
is also evidence for the existence of a (mildly) relativistic outflow associated with the merger \citep{mooley2017, Ruan2017}.

Despite the growing consistency among different theory/simulation results 
-- in favor of mergers, against cc-SNe --
some reservations against the neutron star merger scenario have remained, 
mainly due to the unsettled question whether or not they are consistent 
with the Galactic chemical evolution  \citep{qian00,argast04,matteucci13,hirai2015,shen2015ApJ,vandevoort2015MNRAS,ishimaru2015ApJ,wehmeyer2015MNRAS,Cescutti2015,Cote16}.
The large scatter in Eu abundances at low metallicities compared to that of  $\alpha$-elements is consistent with the picture (where $\alpha$ denotes light nuclei like C, O, and Mg), 
in which $r$-process elements are produced by  phenomena that is much less frequent than normal cc-SNe \citep{Piran14}.
The most likely scenario for explaining the high europium abundances of extremely metal poor stars, is that they were formed in small dwarf galaxies that have merged with the Milky Way \citep{vandevoort2015MNRAS,ishimaru2015ApJ,shen2015ApJ,Naiman2017,Beniamini2017}.
In addition, the chemical abundance studies of dwarf galaxies \citep{tsujimoto2014A&A,ji2016Nature,roederer2016AJ,beniamini+16a,Beniamini2017,Hirai2017} and the abundance measurements of radioactive nuclides of
the solar system materials \citep{paul2001ApJ,turner2004Sci,turner2007E&PSL,brennecka2010Sci,wallner2015NatCo,hotokezakaPP15,tissot2016Sci,tsujimoto2017ApJ} indicate that  rare phenomena as the heavy $r$-process production site.
More recent works focus on the late  chemical evolution in the Milky Way.   The ratio of $r$-process elements to Fe, [Eu/Fe],  declines for [Fe/H]$> -1$,
where ${\rm [X/Y]} = {\rm log} _{10}(N_{\rm X}/N_{\rm Y})- {\rm log} _{10}(N_{\rm X}/N_{\rm Y})_{\odot}$,
$N_{\rm X}$ is the abundance of an element X, and $\odot$ refers to the solar value. It has been questioned whether such a behavior is consistent with the expected merger history in the Milky Way \citep{Cote16,Komiya16}.


Our goal in this article is twofold. First we summarize the cumulative evidences supporting that $r$-process nucleosynthesis takes place  in rare events in which a significant amount of $r$-process elements are produced in each event. 
Using the rate and mass ejection per event inferred from these measurements, we test the neutron star merger scenario for the origin of $r$-process elements in the cosmos. 
 This evidence clearly rules out the normal cc-SNe scenario. Moreover, the rate agrees with merger estimates  from galactic binary neutron stars, from sGRBs, and from GW170817. 
At the same time the amount of matter is consistent with the kilonova/macronova, AT2017gfo,
and  the candidates associated with cosmological sGRBs.
Second, we turn to  the Galactic chemical evolution of $r$-process elements at later times [Fe/H]$\,\gtrsim -1$
and discuss whether the neutron star merger scenario can consistently explain 
the observed distribution of [Eu/Fe]. 

\section{$r$-process production rate,  sGRBs, and GW170817}
\label{sec:rate}

\begin{figure*}[h]
\includegraphics[scale=0.6]{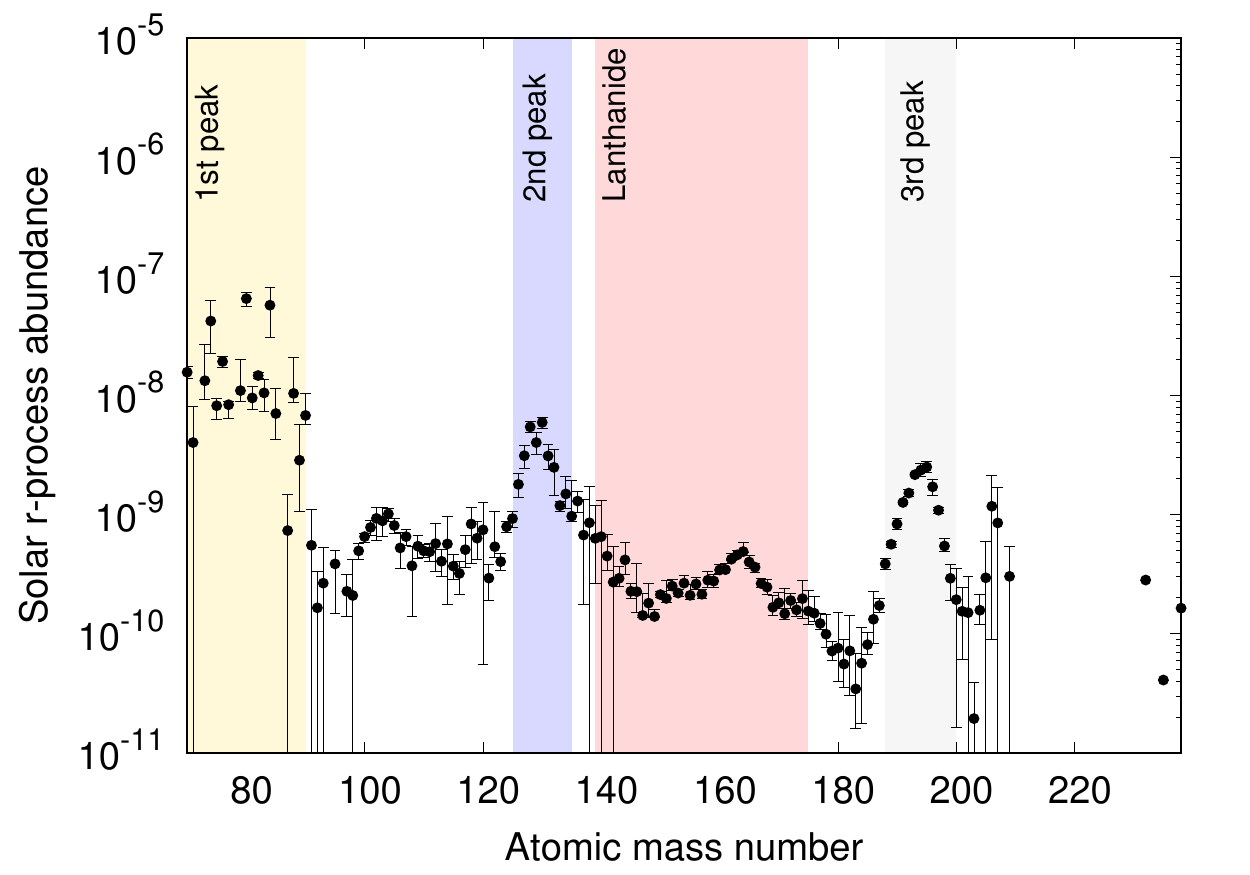}
\includegraphics[scale=0.6]{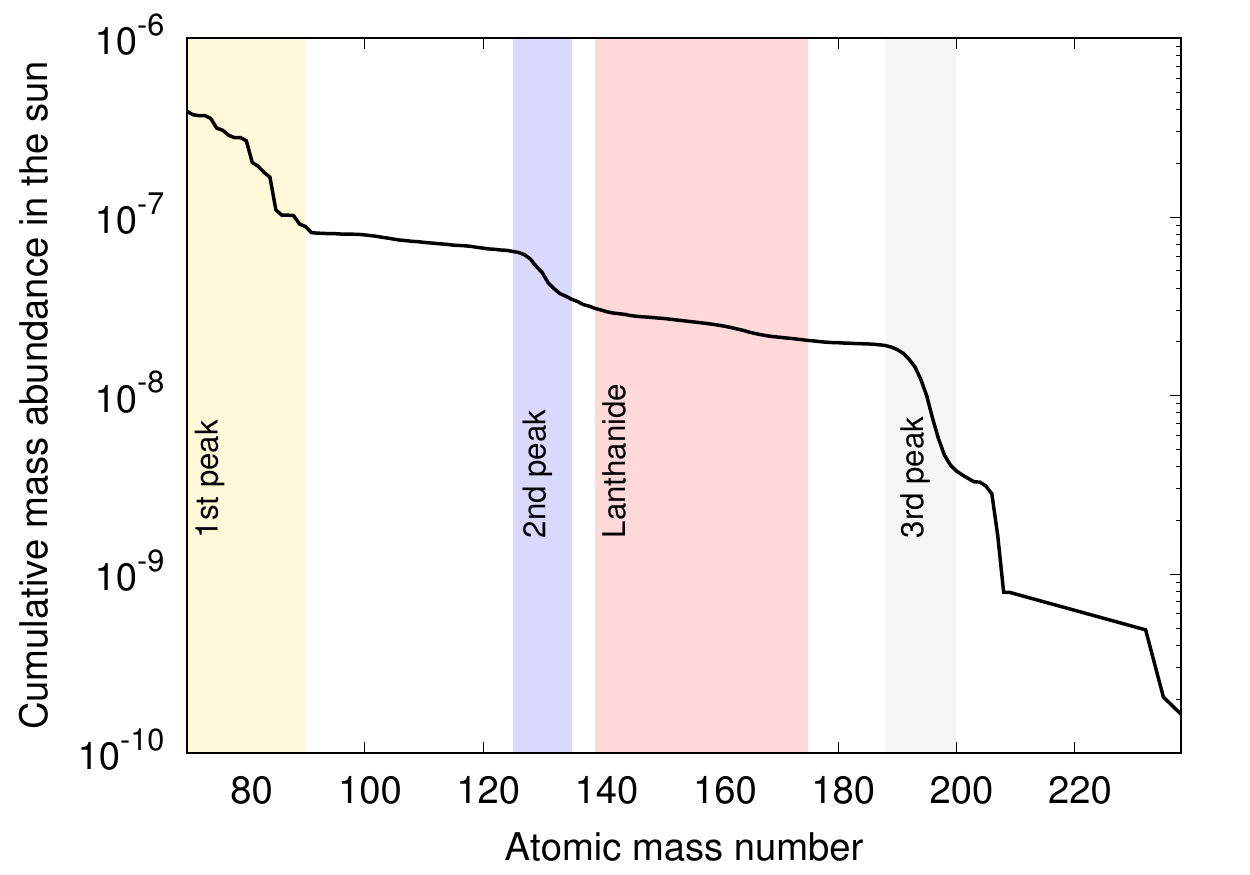}
\caption{The solar abundance pattern of $r$-process elements ({\it left})
and its cumulative abundance ({\it right}). The solar $r$-process abundance pattern is taken from \cite{goriely1999,lodders2003ApJ}.    }
\label{fig:solarabundance}
\end{figure*}

Before discussing details, we describe here the $r$-process abundance pattern. Figure \ref{fig:solarabundance}
shows the solar abundance pattern of $r$-process elements taken from \citep{goriely1999,lodders2003ApJ}.  There are three peaks. For the solar abundance pattern,
most  of the mass of $r$-process elements ($\sim 80\%$) is around the first peak.
However, the abundance ratio of the first peak to the second peak of extreme metal poor stars,
of which the abundance pattern likely reflects a single nucleosynthesis event, 
is often different from that of the solar pattern.  
Some of these stars exhibit abundance patterns beyond the second peak (heavy $r$-process)
that are similar to the solar pattern. However, they don't contain similar amounts of the first peak elements as compared with expectations from the solar abundance pattern (e.g. \citealt{Sneden2008} and 
references therein). 
At the same time, there are  stars that contain a substantial amount of the first peak elements but 
do not show a significant enrichment  of heavy $r$-process elements (e.g. \citealt{Honda2006}). {This suggests that the ratio between ``heavy" and ``light" $r$-process  abundances varies among events or   
there may be different kinds of astrophysical phenomena 
producing ``light" and ``heavy" $r$-process elements.
For instance, electron capture and cc-SNe could produce a sufficient amount of ``light" $r$-process elements (e.g. \citealt{roberts10,wanajo2011ApJ}). }
Therefore, it is important to keep in mind that it is unclear what the minimal atomic mass number of  elements produced by $r$-process events is.  

Since rate estimates of $r$-process events are sensitive to the minimal atomic mass number assumed,
we consider here two scenarios in which an astrophysical phenomenon predominantly produces (i) all the $r$-process elements ($A_{\rm min}=69$) and (ii) 
only heavy $r$-process elements ($A_{\rm min}=90$). The mass fractions of the lanthanides out of the total $r$-process elements for these two cases are $\approx 0.025$ and $0.1$, respectively. In the following, we test the neutron star merger scenario 
for the origin of (heavy) $r$-process elements. 
It is worth noting that numerical simulations of merger ejecta
show that heavy $r$-process elements are robustly synthesized \citep{goriely11a,korobkin12a,Perego2014,wanajo2014ApJ,
eichler2015ApJ,just2015MNRAS,wu2016MNRAS,Siegel2017}. The first peak elements are not necessarily synthesized
and their abundances  depend on the ejection mechanism and on the nature of the remnant  of  mergers
(e.g. \citealt{wanajo2014ApJ,lippuner2015ApJ,wu2016MNRAS,lippuner2017}) so that mergers with different  masses may produce different abundance patterns. 
{Note that there are metal poor stars in which the abundance ratio of Th to Eu is 
larger than that of the sun, the so-called ``actinide boost stars" (e.g. \citealt{Hill2002,Roederer2009}). The variation in elemental abundances beyond
the third $r$-process peak  elements does not affect the mass estimates but they suggest that 
there are $r$-process events that produce larger amounts of very heavy elements. }

\subsection{Production rate  inferred from $r$-process measurements}
\label{sec:tot}
\begin{figure}[h]
\includegraphics[scale=1.2]{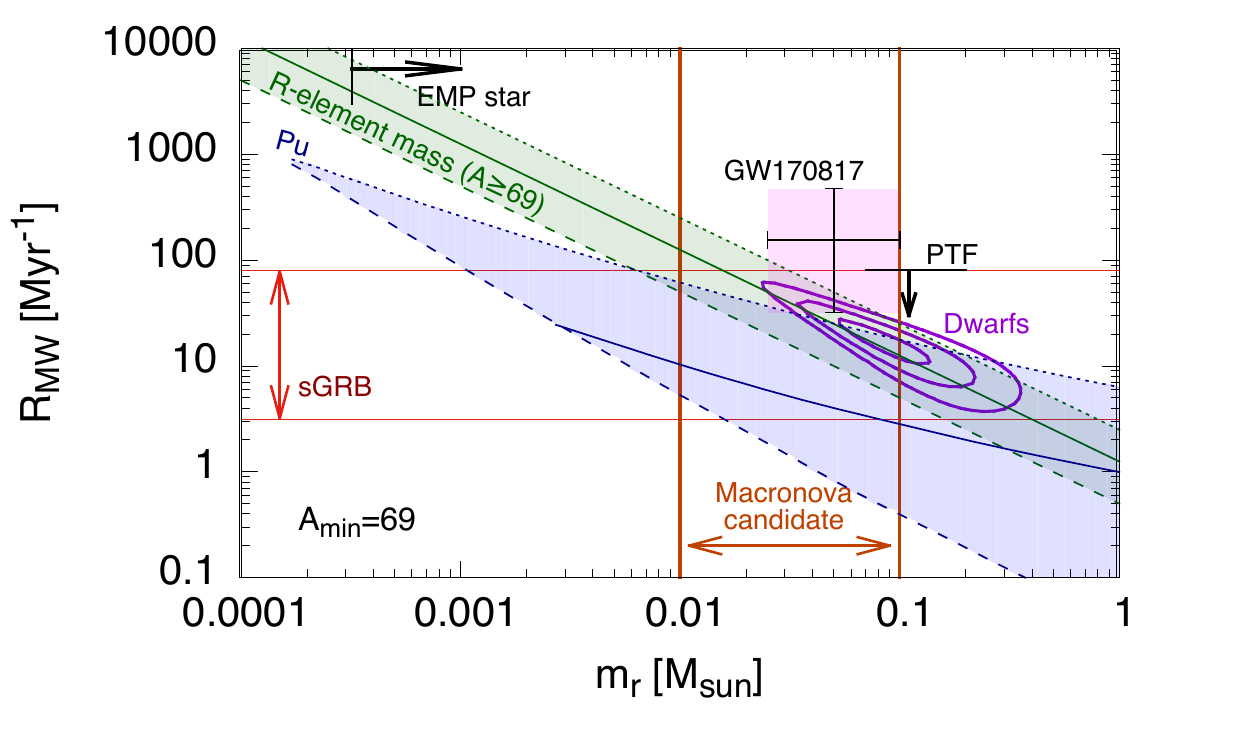}
\caption{The rate of $r$-process production events in the Milky Way and
the  mass produced per event inferred from various measurements (see the text for details and also \citealt{hotokezakaPP15}).
Here, we assume $r$-process elements with the solar abundance 
pattern for $A\geq 69$ (including the first peak $r$-process elements) are produced in each event. 
For GW170817, we take the rate estimated  with the $90\%$ confidence
interval \citep{GW170817} and the mass estimate of $\approx 0.05M_{\odot}$ with an uncertainty of a factor of $2$.
To convert the volumetric event rate ${\rm Gpc^{-3}\,yr^{-1}}$ to the galactic 
event rate, we use the number density of
Milky-Way like galaxies of $\approx 0.01\,{\rm Mpc^{-3}}$. }
\label{fig:consistency}
\end{figure}

\begin{figure}[h]
\includegraphics[scale=1.2]{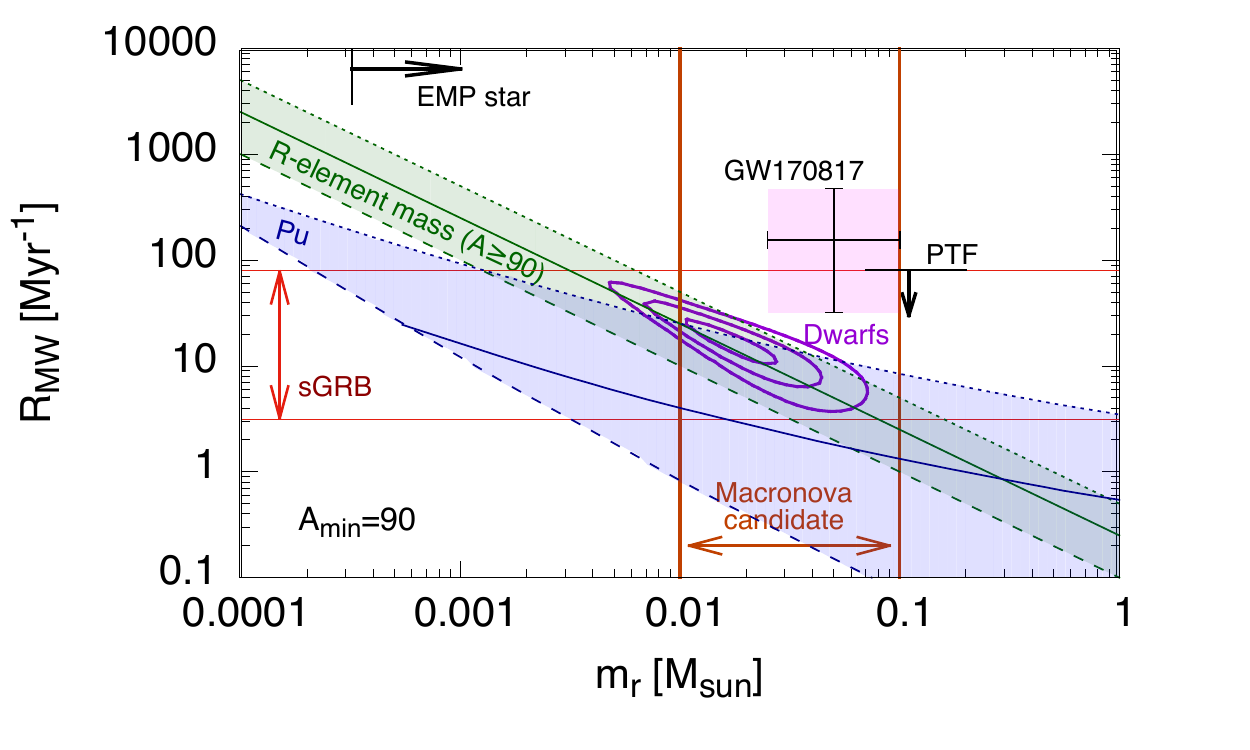}
\caption{Same as Figure \ref{fig:consistency} but for the case that only heavy $r$-process elements 
$A\geq 90$ (excluding the first peak $r$-process elements) are produced in each event. }
\label{fig:consistency2}
\end{figure}

Astrophysical observations and geological 
measurements provide evidence that $r$-process elements are
produced in rare events in each one of those a significant amount of $r$-process elements (a few percent of solar masses) are produced. These observations support the merger scenario.  In this section, we briefly summarize the observations and demonstrate their compatibility with each other and with the merger scenario.  
The results are summarized in Figs. \ref{fig:consistency} and \ref{fig:consistency2}.

{\it (i) The total mass of $r$-process elements in the Milky Way:}
The total mass of $r$-process elements in the Milky Way gives a rough estimate
for the product of  the event rate $R_{\rm_{MW}}$ and the average mass produced by each event
$m_r$: $M_{{\rm tot},\,r}\sim t_{_{\rm MW}}\cdot R_{\rm _{MW}}\cdot m_r$, 
where $t_{_{\rm MW}}\approx 10~{\rm Gyr}$ is the age of the Milky Way (e.g. \citealt{eichler89,Bauswein2014ApJ,hotokezakaPP15,rosswog2017CQG}). This approximate relation assumes that
the mass loss due to the galactic wind is negligible compared to the total stellar mass of the Galactic disk, where we take the total mass of $\approx 6\cdot 10^{10}M_{\odot}$~\citep{mcMillan2011MNRAS}.
Assuming  that the mean $r$-process abundance of stars in the Galactic disk is the solar value
(e.g. \citealt{venn2004AJ,battistini2016A&A} for europium abundances of local stars),
the total mass is estimated as $\sim 23000\,(5000)M_{\odot}$ for elements with $A\geq 69\,(90)$.
This corresponds to 
\begin{eqnarray}
R_{\rm _{MW}}  & \sim  &230\;{\rm Myr^{-1}} (m_r/0.01M_{\odot})^{-1}~~~~ (A_{\rm min}\geq 69),\label{eq:average} \\ 
R_{\rm _{MW}}  &\sim  & 50\;{\rm Myr^{-1}} (m_r/0.01M_{\odot})^{-1} ~~~~(A_{\rm min}\geq 90). \nonumber 
\end{eqnarray}
The estimate involves an uncertainty in the mean abundance 
of $r$-process elements of the Milky Way stars due to the limited 
sample.
Note that this rate is averaged over the Milky Way lifetime so that
the present-day Galactic $r$-process event rate does not necessarily match
this rate and the relation depends  on the redshift distribution of $r$-process production events.
For the neutron star merger scenario, it is  important to note that some neutron star 
mergers do not contribute to the $r$-process enrichment of the 
ISM because they  escape from the
Galactic disk due to the natal kicks (see \S \ref{sec:Kicks}).
This effect may result in an underestimate of  $R_{_{\rm MW}}$.

{\it (ii) Galactic extremely metal-poor stars:} The large dispersion in the [Eu/Fe] abundances of 
extremely metal poor stars compared to that of [$\alpha$/Fe] suggests that  $r$-process production events are rarer than  cc-SNe in which $\alpha$-elements are produced
(e.g. \citealt{Sneden2008,Piran14}). The rate estimate using this distribution depends strongly on the mixing time scale of the ISM and the history of the Milky Way at early times. 
\cite{Macias16} obtained a lower limit on the  mass of $r$-process elements produced per event of  $\sim 0.3\cdot 10^{-3}M_{\odot}$ using the largest value of [Eu/H] obtained from low metallicity stars.  
This estimate is valid as long as $r$-process elements are produced in explosions
with kinetic energy of $\gtrsim 10^{50}$~erg within the metal poor environment. 
Comparing this lower limit with Eq. (\ref{eq:average}) yields $R_{_{\rm MW}} < 1500\; {\rm Myr^{-1}} $, far smaller than 
the galactic cc-SN rate, $\sim 2.8\cdot 10^4 \,{\rm Myr^{-1}} $ \citep{Li2011MNRAS}, but consistent with mergers.

Previous studies have suggested that  metal poor stars in the Galactic halo originate from tidally disrupted dwarf galaxies that merged with the Milky Way \citep{Frebel2010Natur,vandevoort2015MNRAS,ishimaru2015ApJ,Hirai2017}. The low event rate of $r$-process events (compared with SNe) implies that only a small fraction of those dwarfs would have  a single $r$-process event in their history. However, in those that do, the $r$-process enrichment could become extremely large. Indeed, assuming that this is the origin of Galactic halo metal poor stars, \cite{Beniamini2017} have shown that the resulting abundance is consistent with both the mean value and the fluctuations of [Eu/Fe] in halo metal poor stars, supporting the hypothesis that UFDs are the dominant ``building blocks" of the Galactic halo.

{\it (iii) Radioactive elements in the early solar system and in deep sea floor:}
The abundance of radioactive elements in the ISM at a given location 
is determined by their local production history on the time scale of their mean lives
rather than the whole production history. Therefore, the abundance measurements of radioactive
elements allow us to break the degeneracy between $R_{\rm _{MW}}$ and $m_r$ that appears
in Eq. (\ref{eq:average}).
There are two kinds of
measurements of radioactive $r$-process elements.
Their abundances during the formation of the solar system can be inferred from those of the daughter  nuclides 
in the early solar system material, e.g., meteorites and ancient rocks \citep{turner2004Sci,turner2007E&PSL,brennecka2010Sci,tissot2016Sci}.
The abundance measurements of $^{129}$I ($t_{1/2}=15.7\,$Myr), 
$^{232}$Th ($14\,$Gyr), $^{235}$U ($704\,$Myr),
$^{238}$U ($4.5\,$Gyr),
$^{244}$Pu ($81\,$Myr),
and $^{247}$Cm ($15.6\,$Myr) can  constrain the  rate of
$r$-process production events in the vicinity of the early solar system (e.g. \citealt{cowan1991PhR,Wasserburg1996}).
In particular, a relatively large abundance ratio of $^{244}$Pu to $^{238}$U
of $0.008$ implies that the time delay between the $r$-process production and
the formation of the early solar material is not much longer than $81\,$Myr.  
On the other hand,  a tiny amount  of $^{129}$I and $^{247}$Cm  implies 
that the time delay
is much longer than $20$ Myr \citep{tissot2016Sci}.  A time delay  $\tau_{\rm delay}\sim 100\,$Myr
consistently explains these abundance measurements.   
This time delay 
can be approximately translated to 
the  production rate in the Milky Way, $R_{\rm_{MW}}$ \citep{hotokezakaPP15}:
\begin{eqnarray}
\tau_{\rm delay} & \approx  120~{\rm Myr}~(R_{\rm_{MW}}/100~{\rm Myr}^{-1})^{-2/5}(\alpha_t/0.1)^{-3/5}
  (v_{t}/7~{\rm km/s})^{-3/5}(H/0.2~{\rm kpc})^{-3/5}, \nonumber \\
\label{eq.2}
\end{eqnarray}
where $v_{t}$ is the typical turbulent velocity of the ISM, $H$ is the scale height of the ISM disk,
and $\alpha_t$ is the mixing length parameter.

Live radioactive elements, e.g.,  $^{60}$Fe and $^{244}$Pu, are currently accreted by the solar system
and accumulate in the Earth's deep sea floor (see \citealt{Ellis1996} for  an earlier study and \citealt{Fitoussi2008,Fry2016,Wallner2016,Schulreich2017} for studies on $^{60}$Fe). 
While $^{60}$Fe is relatively short lived and is likely produced by SNe, $^{244}$Pu is perfectly suited for our purpose.  
Measurements of the current deposition rate averaged over the last $25$ Myr
show that the abundance of $^{244}$Pu in the ISM that accretes now on the solar system  is much lower ($\sim 10^{-2} $) than 
the abundance at the time of the formation of the  solar system \citep{paul2001ApJ,wallner2015NatCo}. 
 Such a large fluctuation in $^{244}$Pu abundances is consistent with a low event rate of $r$-process production.
\cite{hotokezakaPP15} found the range of values or $R_{\rm _{MW}}$ and $m_r$ that are consistent  with both early solar system and current observations (see also \citealt{tsujimoto2017ApJ}). These values  are shown in Figs. \ref{fig:consistency} and \ref{fig:consistency2}. 

We should note that there are unknown factors in this calculation: (i) the ISM mixing parameter, 
(ii)  the penetration efficiency of dust grains containing $^{244}$Pu from the ISM to the Earth's orbit, 
(iii) the local ISM density where the solar system has been traveling during the last $25$ Myr, and
(iv) the star formation history in the solar neighborhood.  The effect of these factors on the rate and ejecta mass estimate
is shown in \cite{hotokezakaPP15}.

\begin{figure}[h]
\includegraphics[angle = 270, bb=50 50 500 700,width=78mm]{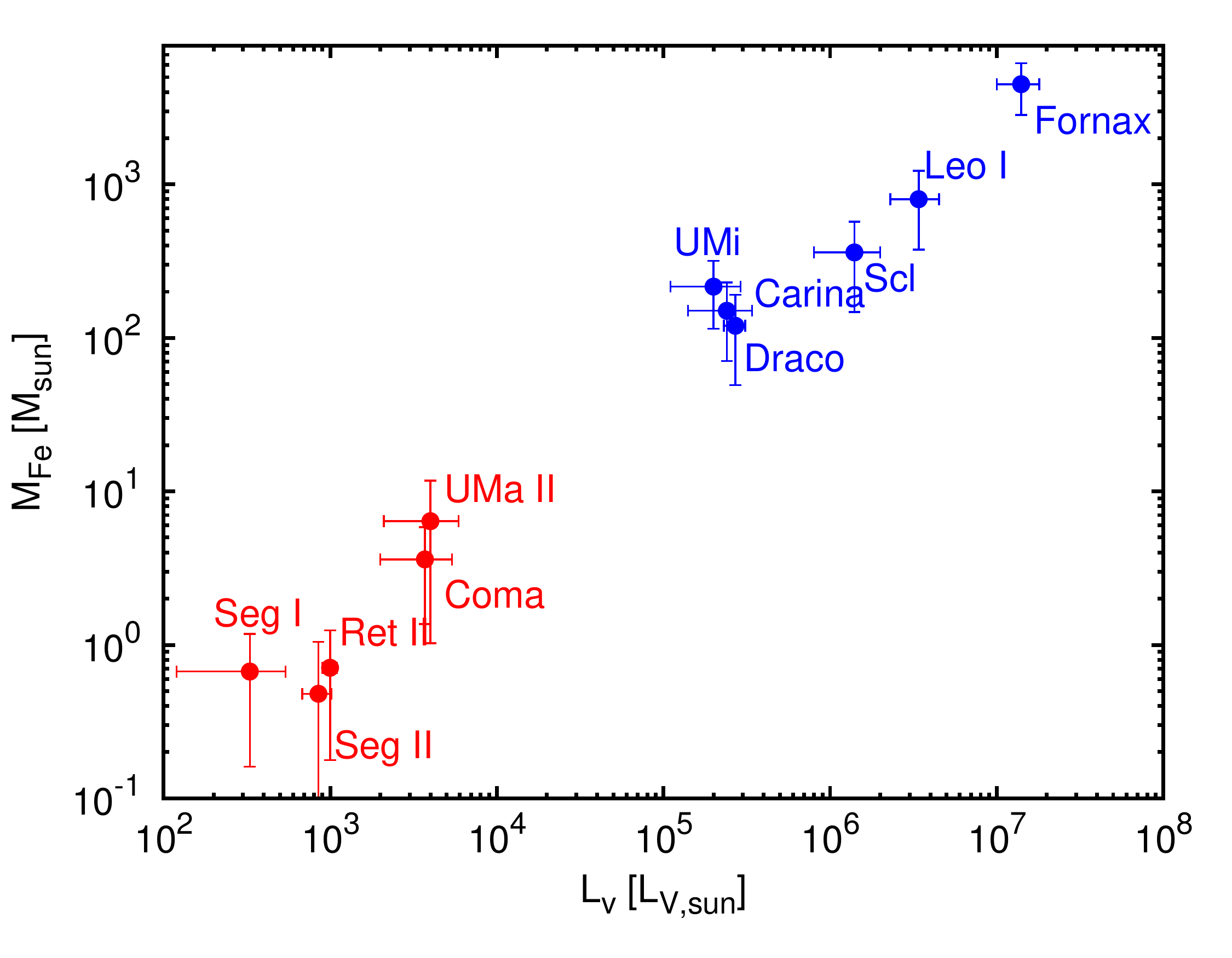}
\includegraphics[angle = 270, bb=50 50 500 700,width=78mm]{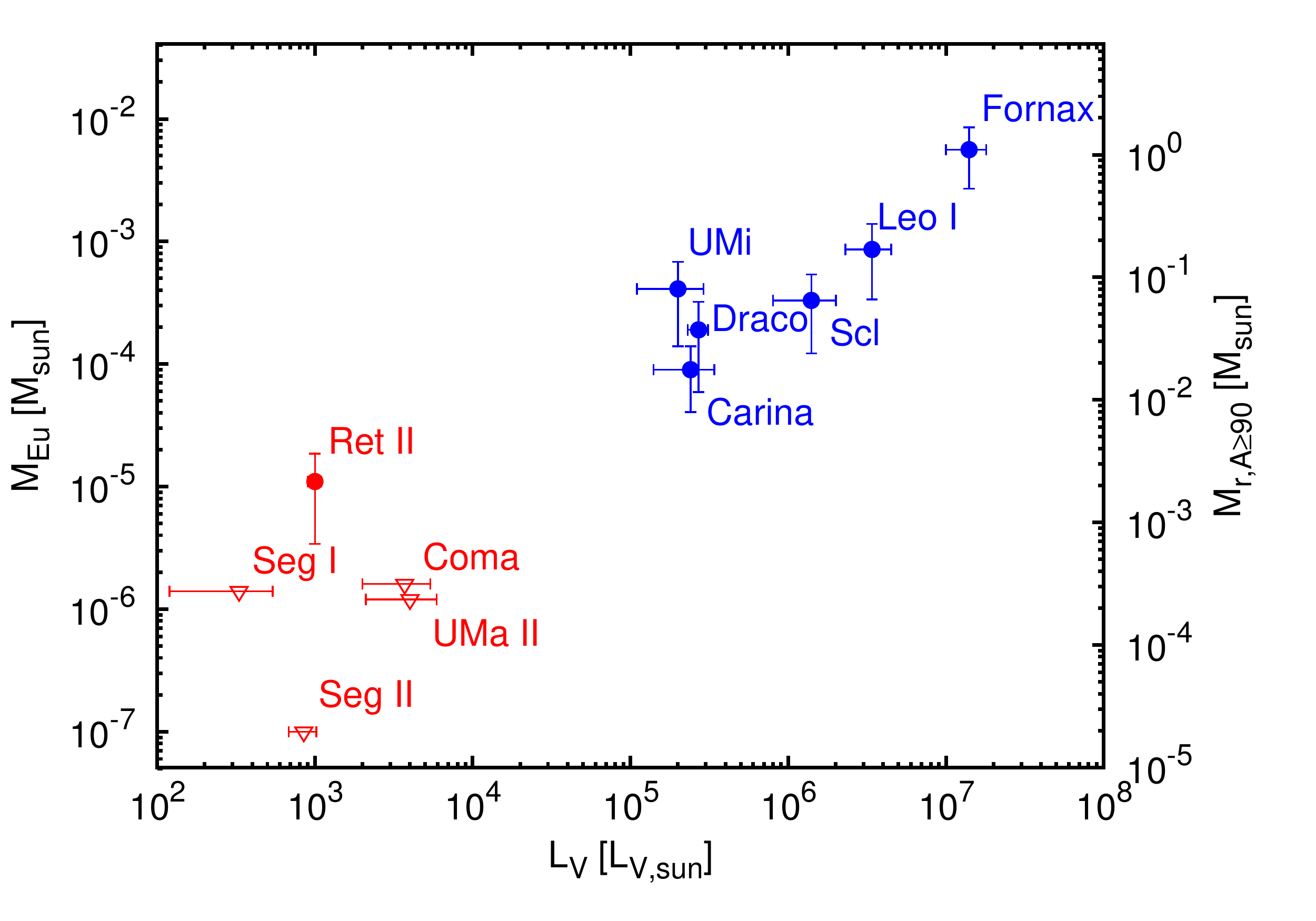}
\caption{
 Iron~({\it left}) and europium~({\it right})  masses produced in dwarf galaxies. The horizontal axis
shows the total V-band luminosity of each galaxy. 
Blue and red points depict the classical dwarfs and UFDs respectively. Open triangles 
show upper limits on the masses. The figures are taken from \cite{beniamini+16a}. }
\label{fig:eu}
\end{figure}

{\it (iv) $r$-process abundances of dwarf galaxies:}
The abundance of $r$-process elements of  dwarf satellite galaxies of the Milky Way 
provides  another estimate. The mean value of [Eu/H] of the observed stars  of classical dwarf galaxies 
is between $-1.5$ to $-0.5$ \citep{shetrone2001ApJ,shetrone2003AJ,venn2004AJ,letarte2010A&A,tsujimoto2015PASJ}. Among ultra-faint dwarf~(UFD)
galaxies,  \cite{ji2016Nature} discovered $r$-process enrichment  in Reticulum 2 with a mean
of [Eu/H] of  $\sim -1$ (see also \citealt{ji2016ApJ,roederer2016AJ}).  
An $r$-process enriched star was recently found in another UFD, Tucana III \citep{hansen2017ApJ}. 
However, only upper limits of $\lesssim -2$
have been obtained for other ultra-faint dwarfs \citep{frebel2010ApJ,frebel2014ApJ,roederer2014MNRAS}. 
As these ultra-faint dwarfs contain several  hundreds to ten thousands stars, these observations 
 suggest that the $r$-process production rate is rarer than one event per $10^4$ stars formed
 (see Fig. \ref{fig:eu}). Using a maximum likelihood analysis, \cite{beniamini+16a} estimated $R_{_{\rm MW}}$ and $m_r$ that are consistent with the  $r$-process elements distribution of  dwarf galaxies. They find   that the rate is  $\approx 1$ per $10^{3}$ cc-SNe,  
coresponding to $R_{\rm _{MW}} \approx 20\; {\rm Myr^{-1}} $  (using a Galactic cc-SNe rate of $\approx 2.8$ SNe per century  \citep{Li2011MNRAS}).

The intersection of this range with the range given by Eq. (\ref{eq:average}) is at the same range of values given by the $^{244}$Pu estimate.
Figures \ref{fig:consistency}  and \ref{fig:consistency2} demonstrate a remarkable consistency between different and independent estimates of the rate of $r$-process nucleosynthesis events and the amount of matter produced in each event. The values show clearly that regular cc-SNe can be ruled out as the sites of (heavy) $r$-process nucleosynthesis. 

Note that \cite{beniamini+16a} assumed that the $r$-process elements produced are fully mixed with
the ISM of which the total mass is proportional to the dark-matter halo mass within the half-light radius of the galaxy.
However, it is still unclear whether this assumption is valid. If the mixing is less efficient, the estimated ejecta mass
is smaller. On the other hand, the estimated ejecta mass is larger, if a significant fraction of
the ejecta is lost from the galaxy.
This issue has been addressed in a one zone analytical model \citep{Beniamini2017}, which suggests that these considerations do not strongly affect the rate and ejecta mass estimates (by a factor of $\lesssim 2$). Numerical studies of the evolution of UFDs (e.g. \citealt{greif2010ApJ,hirai2015,safarzadeh2017}) would be able to consider these issues in further detail. For the neutron merger scenario, other caveats in this estimate are (i) 
it is easier for neutron star binaries to escape from smaller dwarfs  and (ii)
to take place in UFD, mergers must occur within relatively short time scales $\lesssim 1$\,Gyr
as UFDs have quenched their star formation $\approx10\,$Gyr ago.
These effects may result in an underestimate of the merger rate by a factor of a few to $\sim 20$ (e.g. 
\citealt{Beniamini2016a}  and see \S \ref{sec:Kicks}).

While the measurements described above indicate that
(heavy) $r$-process elements are produced in rare events and support
the neutron star merger scenario, this does not rule out rare types of
core collapse events as the source of $r$-process, e.g., strongly magnetized
neutron star formation and black hole formation \citep{suzuki2005ApJ,wanajo2012ApJ,winteler12a,vlasov2014MNRAS,nishimura2015ApJ,fujibayashi2015ApJ,Mosta2017,Vlasov2017}. However, apart from speculating about this possibility,  it is not clear  what is  the expected rate and the amount of $r$-process elements produced in those phenomena.

\vspace{1cm}

\subsection{Short GRB rate and kilonova/macronova candidates}

Neutron star mergers are considered as the classical progenitors of 
sGRBs \citep{eichler89}. Information on their rate and possible mass ejection that takes place in them can shed additional light on the problem at hand. 

{\it (i) Rate}: Observed short GRBs occur typically at a redshift of $\approx 0.5$, which is much further 
than GW170817. 
\cite{WP15}   find  $R_{\rm SGRB} = 6 \pm 2\;{\rm Gpc}^{-3} {\rm yr}^{-1}$ which agrees within a factor of 2 with various previous estimates \citep{guetta06,NGF06,GS09}.
Assuming  an unknown beaming correction  $f_b= (1-\cos\theta_j)$ to be  $100$ (e.g. \citealt{fong13b}),
the rate is $600\,{\rm Gpc^{-3}\,yr^{-1}} (f_b^{-1}/100)$ and the Galactic rate is $\approx 60\;{\rm Myr}^{-1} (f_b^{-1}/100)$, using the density of galaxies of $0.01\,{\rm Mpc^{-3}}$.
This estimate depends on the luminosity function of sGRBs and particularly on the lowest peak luminosity 
of sGRBs.  \cite{WP15} adopt  $5\cdot 10^{49}$ erg/s,
with which the {\it Swift} sGRB rate does not exceed the  BATSE rate.
Note that the luminosity of the $\gamma$-ray pulse associated with GW170817 is much lower than
this value. However, the lack of the X-ray afterglow  until $9$ days after the merger suggests  
(if there was one) did not point towards Earth  \citep{Troja2017,Evans2017,Haggard2017,Margutti2017}.  Therefore, we should keep in mind that this $\gamma$-ray burst, GRB 170817A, is not the typical 
sGRB but a new type of GRBs,  a {\it low-luminoisty} short GRB.

{\it (ii) Time delay}: The population of  sGRB hosts shows evidence for a time delay between the burst and the progenitor formation. 
The most notable one is that a fraction of bursts ($\sim 20\%$) take place in early-type galaxies,
where  star formation is no longer active (e.g. \citealt{Berger14}). A comparison of  the redshift distribution of
sGRBs to the cosmic star formation history, \cite{WP15} and \cite{Ghirlanda2016} show that there is a time delay between the two.  A  delay-time distribution  $\propto t^{-1}$ with a minimal time delay of $20$\,Myr is consistent with the observed rate. However, these parameters, the slope of the time delay distribution and the minimal delay time  are rather weakly constrained \citep{WP15}. 

{\it (iii) Ejecta mass estimate:} 
Observations of optical-nIR signatures that follow sGRBs and can be considered as kilonova/macronova candidates, GRB 080503 (\citealt{perley09}; but its redshift is unknown),  
GRB130603B \citep{Tanvir13, Berger13}, GRB 060614 \citep{Gal-Yam2006,Yang+15,Jin2015}, GRB 050709 \citep{Fox2005,Watson2006,Jin+16}, and GRB160821B \citep{kasliwal2017ApJ}, provide  estimates of the amount of matter ejected in mergers, $m_r$, from the observed light curves. 
The estimated values are $0.01 \lesssim m_r \lesssim 0.1 
M_\odot$ \citep{Tanvir13, Berger13,hotokezaka2013ApJ,kasen2015MNRAS,Yang+15,Jin+16,kasliwal2017ApJ}. These values are shown 
in Fig. \ref{fig:consistency}.  Therefore, if the kilonova/macronova interpretation is correct (all current kilonovae/macronovae identifications, apart from AT2017gfo, are based on just a single or at most a few data points),   then when  combined with the rate estimates, mergers are the dominant source of heavy $r$-process material in the universe.  Note that the optical-nIR bumps of GRB 050709, 130603B, and 170821B are roughly consistent with the photometric light curve of AT2017gfo.  The optical bump of GRB 060614 is much brighter than  AT2017gfo. Such a bright signal requires a larger ejecta mass and may be associated with a black hole-neutron star merger \citep{Yang+15,Jin2015}.

{The radio counterparts to kilonovae/macronovae are expected 
to arise on time scales from weeks to years \citep{NP11}. The 
follow-up radio observations for sGRBs have not detected
late-time radio flares \citep{metzger13,Horesh2016,Fong2016}.
These upper limits constrain the kinetic energy of the merger outflows
$\lesssim 10^{52}$--$10^{53}$ erg, which is much larger than the ones expected for dynamical ejecta but rules out the existence of a  magnetar central engine.}

{\it (iv) Evidence for mass surrounding sGRBs:}
Moharana and Piran (2017)\citep{moharana2017} have recently shown that the duration distribution of sGRBs shows a plateau at short durations (less than $0.4$ s). This can be interpreted as the effect of a jet penetration through an engulfing envelope of surrounding matter. Scaling from corresponding estimates for LGRBs \citep{Bromberg2011} they find $ t_{\rm break} = 0.4 (L/10^{49} {\rm erg})^{-1/3} (\theta/ {10^{\circ}})^{4/3}( R/10^9{\rm cm})^{2/3} (m_r/{0.1 M_\odot})^{1/3}$  s, where  $L$ and  $\theta$ are the energy and opening angle of the jet and $m_r$ and $R$ are mass and size of the surrounding material. This results suggests the existence of ejecta with $\sim 0.1 M_\odot$ surrounding the merger.

\subsection{GW170817 and its electromagnetic counterparts}

On 17th of August in 2017,  a gravitational-wave signal from 
a binary neutron star merger was detected by the advanced LIGO/Virgo detectors \citep{GW170817}. 
The total mass and mass ratio of the binary are determined as $2.74^{+0.04}_{-0.01}M_{\odot}$ ($2.82^{+0.47}_{-0.09}M_{\odot}$)
and 0.7--1.0 (0.4--1.0)
for the $|\chi|\leq0.05$ ($0.89$) prior \citep{GW170817}. The total mass is similar to those measured in Galactic binary neutron stars (e.g. \citealt{Ozel2012,Tauris2017}).
The gravitational-wave luminosity distance to the source is  $40^{+8}_{-14}$\,Mpc. 
About two seconds later, a weak short-duration $\gamma$-ray pulse, GRB 170817A, was found   
by the {\it Fermi} GBM \citep{gammaGW170817,Goldstein2017}.
A $\gamma$-ray signal was also found in the data of the {\it INTEGRAL} satellite \citep{Savchenko2017}.  
The localization area of GRB 170817A is consistent with the area determined by LIGO and Virgo. 
Although  GRB 170817A is extremely weak compared to sGRBs (the isotropic 
equivalent $\gamma$-ray energy is  smaller  than the weakest sGRB by more than two orders of
magnitude), the observed $\gamma$-rays indicated that a relativistic outflow was   produced
 \citep{Kasliwal2017,Gottlieb2017}. An optical-nIR counterpart (AT2017gfo) was discovered at $\sim 11$\,hr after the merger \citep{Swope, Soares-Santos2017, Valenti2017, Arcavi2017Natur, Tanvir2017, Lipunov2017}. This enabled the identification of the host galaxy NGC 4993, which is 
located at a distance consistent with  $40^{+8}_{-14}$\,Mpc \citep{Hjorth2017}.
 Numerous follow-up observations were conducted and the photometric light curves \citep{Andreoni2017,Arcavi2017Natur,Cowperthwaite2017,Evans2017,Lipunov2017,Kasliwal2017,Kilpatrick2017,Smartt2017,Valenti2017,Drout2017,Pian2017,Swope,Tanvir2017,Utsumi2017} and
 spectral evolution \citep{Andreoni2017,Buckley2017,Chornock2017,Kilpatrick2017,Nicholl2017,Shappee2017,Smartt2017,Kasliwal2017,McCully2017,Pian2017,Tanvir2017,Troja2017,Valenti2017,abbott2017ApJ}
 are broadly consistent with a kilonova/macronova model (see, e.g., \citealt{Kasen2017,Tanaka2017} for  modelings).
At later times, an X-ray counterpart was discovered by the {\it Chandra X-ray Observatory} \citep{Troja2017} and a radio counterpart was discovered by the Karl G. Jansky Very Large Array  \citep{Hallinan2017} 
at the same location of AT2017gfo (see also \citealt{Evans2017,Haggard2017,Margutti2017,Ruan2017} for  the X-ray observations and \citealt{Alexander2017,Kim2017,mooley2017} for the radio observations). These emissions are most likely
arising from the forward shock between the ejected outflow and ISM (see, e.g., \citealt{Kasliwal2017} for an interpretation of the multi-wavelength observations).

{\it (i) Rate and time delay:} The rate of binary neutron star mergers as implied from GW170817
is estimated as $1540^{+3200}_{-1220}\,{\rm Gpc^{-3}\,yr^{-1}}$.  
This can be compared with  estimates  from the observed population of binary pulsars
\citep{NPS91,Phinney91}. Recent estimates \citep{abadie10} suggest $\approx 1$ -- $10^3\;{\rm Myr}^{-1}$,
which is broadly consistent with GW170817 (see also \citealt{kim2015MNRAS}). 
Given this rate and the estimate
of the ejected $r$-process mass for AT2017gfo (see below), neutron star mergers could
predominantly produce $r$-process elements in the Milky Way (Figs. \ref{fig:consistency} and \ref{fig:consistency2} and see also 
\citealt{Cote2017,rosswog2017}).   
It is worth noting that the bright optical signal of
AT2017gfo and the estimated rate suggest that  optical transients similar to AT2017gfo could have
been observed by optical transient surveys. However, Kasliwal {\em et.~al.} (2017)\citep{Kasliwal2017} did not
 found such objects in the database of the Palomar Transient Factory, indicating
that the rate is $\lesssim 800\ {\rm Gpc^{-3}\,yr^{-1}}$. 

The host galaxy properties have an implication on the time delay between
the merger and progenitor formation.
NGC 4993 is an S0-type galaxy and its current star formation rate, averaged over the
last 100 Myr, 
is quite low $\sim 0.01 M_{\odot}\,{\rm yr^{-1}}$\citep{Blanchard2017}.  \cite{levan2017} have shown that the galaxy is dominated by an old
stellar population and a very small fraction of  stars ($\ll 1\%$ of the stellar mass) 
were formed recently $\lesssim 500$~Myr (see also \citealt{Blanchard2017} and \citealt{Pan2017}).  In addition, there is no young stellar mass
system around the location of AT2017gfo \citep{levan2017}.
 Therefore, GW170817 is likely to have  a merger time of $\sim 1$--$10$~Gyr
  \citep{Blanchard2017,levan2017,Pan2017}. 
Finally, the projected distance of the merger of $2$\,kpc away from the center of the host galaxy indicates that 
the system did not receive a strong natal kick of $\gtrsim 200$\,km/s \citep{Blanchard2017,levan2017,Pan2017,Abbott2017kick}.

{\it (ii) Ejecta mass and composition:}
The kilonova/macronova light curves approximately satisfy the following relations for a given 
ejecta mass, velocity, and opacity. The light curve rises on a diffusion timescale:
\begin{eqnarray}
t_{\rm p} \approx \sqrt{\frac{\xi \kappa M_{\rm ej}}{4\pi cv_{\rm ej}}}\approx 5\,{\rm days}\,\xi^{1/2} \left(\frac{\kappa}{10{\rm \,cm^2/g}}\right)^{1/2}
\left(\frac{M_{\rm ej}}{0.01M_{\odot}}\right)^{1/2}\left(\frac{v_{\rm ej}}{0.1c}\right)^{-1/2},
\end{eqnarray}
where $\kappa$ and $v_{\rm ej}$ are the ejecta's opacity and velocity, and $\xi$ is an order unity parameter depending on the ejecta's
density profile.
The luminosity and the effective temperature are given as
\begin{eqnarray}
L_{\rm bol} (t>t_{\rm p})  \approx  M_{\rm ej}\cdot \dot{Q}(t)&\approx& 2.5\cdot 10^{40}\,{\rm erg/s}\,\left(\frac{t_p}{5\,{\rm day}} \right)^{-1.3}\left(\frac{M_{\rm ej}}{0.01M_{\odot}}\right),\\
T_{\rm eff} (t_{\rm p}) \approx  \left(\frac{L_{\rm bol}(t_{\rm p})}{4\pi \sigma v_{\rm ej}^2 t_{\rm p}^2} \right)^{1/4} &\approx& 2200\,{\rm K}\, \left(\frac{L_{\rm bol,p}}{2.5\cdot 10^{40}\,{\rm erg/s}} \right)^{1/4}
\left(\frac{v_{\rm ej}}{0.1c} \right)^{-1/2}
\left(\frac{t_{\rm p}}{5\,{\rm day}} \right)^{-1/2},
\end{eqnarray}  
where the $\dot{Q}$ is the radioactive heating rate per unit mass and $\sigma$ is the Stefan-Boltzmann constant.
Here, we use $\dot{Q}(t)\approx 10^{10}t_{\rm day}^{-1.3}$ erg/s/g, which is a general feature of
$\beta$ decay heating rate resulted from a statistical ensemble of many $r$-process decay chains (e.g. \citealt{metzger10b,korobkin12a,Hotokezaka2017}). 
 Note that the specific heating rate depends on
  the thermalization efficiency, which varies with time because of the  $\gamma$-ray escape, the inefficiency of the electrons' thermalization, and the contribution of $\alpha$-decay and spontaneous fission \citep{Hotokezaka2016MNRAS,barnes2016ApJ}.

One of the most important features of kilonovae/macronovae is a high opacity $\kappa\approx 10\,{\rm cm^2/g}$ 
when lanthanide elements substantially exist in the ejecta \citep{barnes13a,kasen13a,tanaka13a,Tanaka2017arXiv}. This high opacity delays the peak time and
the color at the peak becomes redder (red kilonova/macronova).
  In the case that lanthanides are absent, the opacity is rather low
$\kappa\approx 0.1$--$1\,{\rm cm^2/g}$ \citep{kasen2015MNRAS,Tanaka2017arXiv}
similar to that of the iron group (blue kilonova/macronova).

AT2017gfo rises to $\approx 10^{42}\,{\rm erg/s}$ on a time scale of $\approx 0.5$\,day
and the early spectrum at $\approx 1.5$ days can be broadly described a  thermal spectrum
with $T\approx 5000$\,K (e.g. \citealt{Pian2017,Smartt2017}). 
This requires  an ejecta component of $\sim 0.01M_{\odot}$
that does not contain a significant amount of lanthanides.
At a few days after the merger, the bolometric light curve declines approximately as $\approx t^{-1.3}$, then it starts to decline steeper around $6$ days 
(e.g. \citealt{Cowperthwaite2017,Drout2017,Kasliwal2017,Waxman2017}).
The late spectral structure is broadly consistent with a kilonova/macronova model of which
the mass fraction of lanthanides is $\approx 10^{-3}$ --$10^{-2}$ and for which 
 another ejecta component with $\sim 0.04M_{\odot}$ is required 
(\citealt{Kasen2017,Tanaka2017,Metzger2017arXiv,Waxman2017,Perego2017,Villar2017}).
Although specific atoms are not identified in the optical-nIR signal,
the composition of the ejecta that can consistently explain the observed signal indicates the following:
\begin{itemize}
\item The early emission requires a component of $\sim 0.01M_{\odot}$ without a significant
contamination of lanthanides, i.e., $A<140$.
\item The late  emission requires a component of $\sim 0.04M_{\odot}$
containing lanthanides (the mass fraction of $\sim 10^{-3}$--$0.01$).  
\item The decline rate of the bolometric light curve is consistent with that expected from 
the heating rate resulting from an assembly of numerous $\beta$-decay chains.
\end{itemize}
There are uncertainties in the mass estimates by a factor of a few due to
the unknown composition and profile of the ejecta. It is worth noting that
the  mass fraction of lanthanides out of the total $r$-process elements 
of $\approx 10^{-3}$--$10^{-2}$
 is smaller than the expectation from
 the solar matter with $A_{\rm min}=90$ by a factor of $\approx 10$--$100$
 and it is rather consistent with the one for $A_{\rm min}=69$, of which the mass fraction is $\sim 0.01$.
 This implies that a significant amount of 
 the first peak  $r$-process elements may be synthesized in the ejecta. 
{Note however that lanthanide-rich materials moving at slow velocities may not be seen in the optical-nIR data. Therefore, it is possible that the actual mass fraction of lanthanides is higher than that indicated by the observed data. }

Overall there is a remarkable consistency between the rates of  sGRBs and GW170817, the mass  ejection observed in kilonova/macronova candidates and in AT2017gfo, and the estimates of $R_{\rm _{MW}}$ and $m_r$ from $r$-process elements observations.  This clearly motivates us to turn to the question addressed now - whether mergers are consistent with the observed chemical history of the Milky Way. However, before turning to that we discuss the delay time of mergers and the natal kicks.

\section{Merger time and Natal Kicks}
\label{sec:Kicks}

The merger time and the natal kicks have an important effect on both the total amount of $r$-process nucleosynthesis and on the temporal evolution. 

{\it (i) Merger time distribution:} The orbits of binary neutron stars decay due to gravitational-wave radiation. 
The time until merger, $\tau_{\rm GW}$, is related to the initial semi-major axis, $a$, and eccentricity, $e$, as
$\tau_{\rm GW}\propto a^4 (1-e^2)^{7/2}$.  The merger time is very sensitive to the initial semi-major
axis, e.g., a difference in $a$ by a factor of $4$ corresponds to a  different 
merger time by a factor of $256$. Given an initial semi-major axis distribution, $dN/da\propto a^{-\alpha}$,
the delay-time distribution is $dN/d\tau_{GW}\propto \tau_{\rm GW}^{-(\alpha-3)/4}$. Therefore,
in order to get a somewhat steep delay-time distribution, the initial semi-major axis distribution
must be very steep.

 Observationally, neutron star binaries exhibit a range of delay times between their formation and mergers. For instance, 
 the galactic binary neutron stars  have merger times distributed 
in different time scales from $<100$~Myr to  longer than a Hubble time
\citep{Lorimer2008}. 
Furthermore, 
the redshift distribution of sGRBs also shows that 
sGRBs occur with the delay time distribution of $t^{-0.8^{+0.25}_{-0.24}}$ from
the cosmic star formation rate \citep{WP15}\footnote{The upper and lower values show the $1\sigma$ uncertain interval.
Note that the mean and $\sigma$ values depend on the shape of the cosmic star formation history.}. Finally,  GW170817 occurred in an S0-type galaxy of which the star formation rate is very low
and the time delay between the double neutron star formation and the merger 
is likely $\gtrsim 1$\,Gyr  \citep{Blanchard2017,levan2017,Pan2017}. 

{\it (ii) Kick:} A crucial ingredient in the formation of binary neutron star systems is the kick imposed on the center of mass during the stellar collapse that leads to the formation of the second born neutron star. Two factors contribute to the change in the center of mass velocity. The first is simply  the amount of mass ejected during the collapse, $\Delta M$.   The second is the asymmetry in the collapse, and it is proportional to $\vec{v}_k$, the kick velocity imparted to the newly born neutron star by the ejected mass. The distributions of $\Delta M$ and $\vec{v}_k$ have been studied both directly from observations of Galactic binary neutron stars' orbital parameters \citep{Fryer1997,Wex2000,vdH2004,Piran2005,Thorsett2005,Stairs2006,Kalogera2007,Ferdman2014,Dall'Osso2014,BP2016,Tauris2017}
and in combination with population synthesis models \citep{Podsiadlowski2004,Wang2006,Willems2006,Belczynski2008,Wong2010,Schwab2010}. The emerging evidence points towards a bimodal population of stellar collapses in those systems. The second collapse in the majority of systems involves small amounts of mass ejection $\Delta M\lesssim 0.5M_{\odot}$ and weak kicks $v_k \lesssim 30 \,\mbox{km s}^{-1}$. A neutron star formation scenario of this kind has been suggested based on various theoretical grounds \citep{Miyaji1980,Nomoto1987,DewiPols2003,Ivanova2003,Podsiadlowski2005,Tauris2015}. Only a minority of the systems have formed via a standard supernova involving larger mass ejection
of $\sim 2.2 M_{\odot}$ and kick velocities of up to $400$\,km\,s$^{-1}$. The observed orbital parameters of binary neutron star systems are also consistent with a possible correlation between the kick velocities and mass ejections \citep{BP2016}. Specifically, if the shell always receives the same speed due to the explosion and has a similar degree of asymmetry, then there should be a linear correlation between the kick imparted to the neutron star and the amount of mass ejected. Note that this model is consistent with the existence of the two groups of systems described above.

The distribution of binary neutron star systems' center of mass velocities has important implications for $r$-process formation both in dwarf galaxies and inside our Galaxy. We consider first the former case. In recent years, \cite{ji2016Nature} and \cite{roederer2016AJ} reported on the discovery of an $r$-process enriched UFD galaxy Reticulum II. The total stellar luminosity of Reticulum II is only $\sim 10^3L_{\odot}$ and the velocity dispersion is $\sim 4$~km/s~\citep{walker2015ApJ}. Based on this detection, \cite{Bramante2016} argued against a neutron star merger origin for this material, suggesting that the kick imparted on the system during the formation of the second born neutron star, would have ejected the binary out of such a low mass galaxy. An additional issue is that UFDs are composed of very old stellar material, implying that the binary neutron star should merge within the first $\sim $Gyr of its formation (before star formation ceases), in order to successfully enrich future stars with $r$-process material. Whether or not a significant population of such rapid mergers exist, crucially depends on the initial distribution of binary separations.
\cite{Beniamini2016a} considered the distributions of mass ejections and kicks as implied by Galactic binary neutron star observations. They have shown that a large fraction of  systems would have center of mass velocities $<15\,\mbox{km s}^{-1}$ (comparable to the lowest escape velocities from UFDs). Furthermore, two of the ten binary neutron star systems (that do not reside in globular clusters and for which the orbital parameters can be estimated) have short spin-down times, implying small initial separations and short overall lifetimes. One of these is the double pulsar, which also has a small observed proper motion $\lesssim 10 \,\mbox{km s}^{-1}$ \citep{Kramer2006}, demonstrating that ``rapid mergers" can have slow center of mass velocities. \cite{Beniamini2016a} have used these observations to show that a significant fraction ($0.06\lesssim f\lesssim 0.6$) of binary neutron star systems are expected to both remain confined in UFD galaxies and merge within less than a Gyr.

 Finally, we consider the effects of natal kicks on $r$-process formation in our own Galaxy. The observation of $r$-process elements in stars with values of $\mbox{[Fe/H]}\approx -3$ and below, suggests a significant amount of $r$-process formation in the first few Myr since the Galaxy's formation. This again, has been used as an argument against a neutron star merger origin \citep{argast04}. However, the small initial separations of some binary neutron star systems inferred from observations, as well as the fact that a fraction of systems can increase their eccentricity and/or reduce their separation as a result of kicks, both contribute to a significant population of ``rapid mergers" \citep{Beniamini2016a}. These may help account for the rise in $r$-process abundances at early times. Furthermore, even with a delay of few tens of Myrs between the SNe leading to the formation of the neutron star's in a binary neutron star system and their eventual merger, the binary neutron star can travel a large distance from their birth place. Thus,  mergers can naturally account for the observations of significant amounts of $r$-process material in some low metallicities halo stars in the Galaxy and to the large fluctuations in [Eu/Fe]  in metal poor stars.

\section{Galactic Eu Evolution}

We already noted that the large fluctuations in [Eu/Fe] are consistent with rare events \citep{Piran14} and
the large amounts of heavy $r$-process material observed in some low metallicity stars put a lower limit
on the yield of each event \citep{Macias16}.  
Depending on the origin of these low metallicity halo stars, this may require a short minimal time delay between the star formation and the first merger - an issue to which return later
\citep{argast04,matteucci13,hirai2015,shen2015ApJ,vandevoort2015MNRAS,ishimaru2015ApJ,wehmeyer2015MNRAS,Cescutti2015,Cote16}.
Here, we focus on the late time chemical evolution of [Eu/Fe] of the Milky Way
and consider that the abundances of europium and $\alpha$-elements
trace the $r$-process production and cc-SNe history, respectively. The abundance 
distribution of iron follows a combination of cc-SNe and SNe Ia. 
Remarkably,  [Eu/Fe] and [$\alpha$/Fe] evolve quite similarly (indistinguishable 
within uncertainties) for [Fe/H] $\gtrsim-1$ (Fig. \ref{fig:abundance}).
Roughly speaking, the similarity in the mean value of [$\alpha$/Fe] and [Eu/Fe] means 
that a significant fraction of europium must be produced before the typical time it takes
SNe Ia start to produce iron.  This feature is naturally explained if the $r$-process
sources  immediately follow the Galactic star formation \citep{argast04,matteucci13,ishimaru2015ApJ,Cote16}.

It  has been  claimed that the observed [Eu/Fe] 
distribution for [Fe/H] $\gtrsim -1$ can be reproduced by neutron star mergers
with a very steep delay time distribution or a constant delay time of $\sim 100$~Myr \citep{matteucci13,wehmeyer2015MNRAS,Cote16}. Recently, \cite{Cote16} and \cite{Komiya16}
claimed that this late-time behavior of [Eu/Fe] is inconsistent with the 
delay time distribution $\propto t^{-1}$.  However, neutron star binaries exhibit a range of delay times between their formation and mergers (see \S \ref{sec:Kicks}). In the followings, we discuss whether such a delay time distribution is 
compatible to the galactic distribution of [Eu/Fe] for [Fe/H]$>-1$.
 
Before discussing details of the Galactic chemical evolution,
let us focus on the event rate of core-collapse SNe, SNe Ia, and mergers,
and the total amount of iron and $r$-process elements produced by these events.
We define a simple form of the delay time distribution as
\begin{equation}
D_i(t) = \frac{C_{i}\theta(t-t_{\rm min})}{t^{b_i}},
\end{equation}
where $t$ is the delay time between a SN Ia or a merger and the time when
the progenitor is formed, $t_{\rm min}$ is the minimal delay time,
$C_i$ is a normalization factor, which is calibrated
by the local event rates. For  SNe Ia,
we take $b_i=1$ as inferred from observations (e.g. \citealt{Totani2008,Maoz2014}).

\begin{figure*}[h]
\includegraphics[scale=0.6]{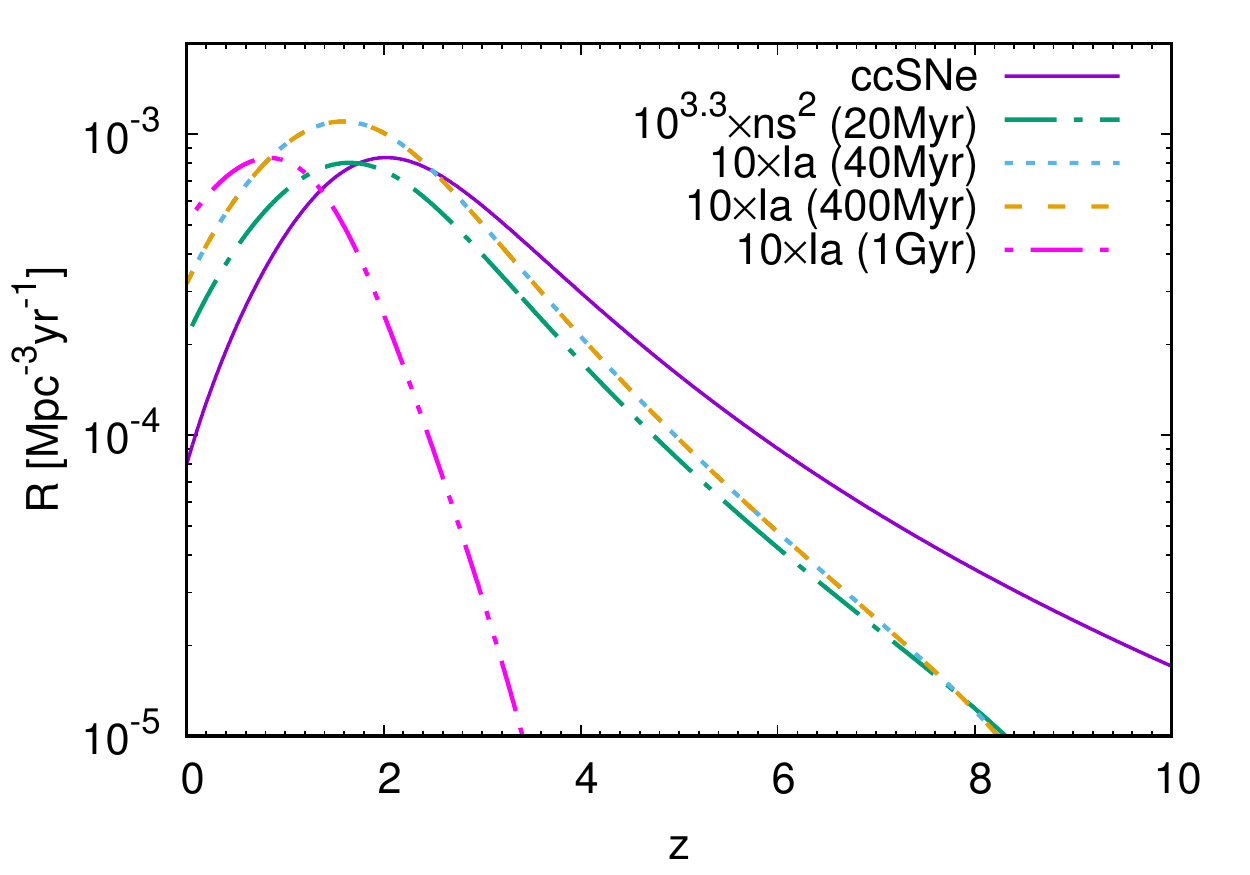}
\includegraphics[scale=0.6]{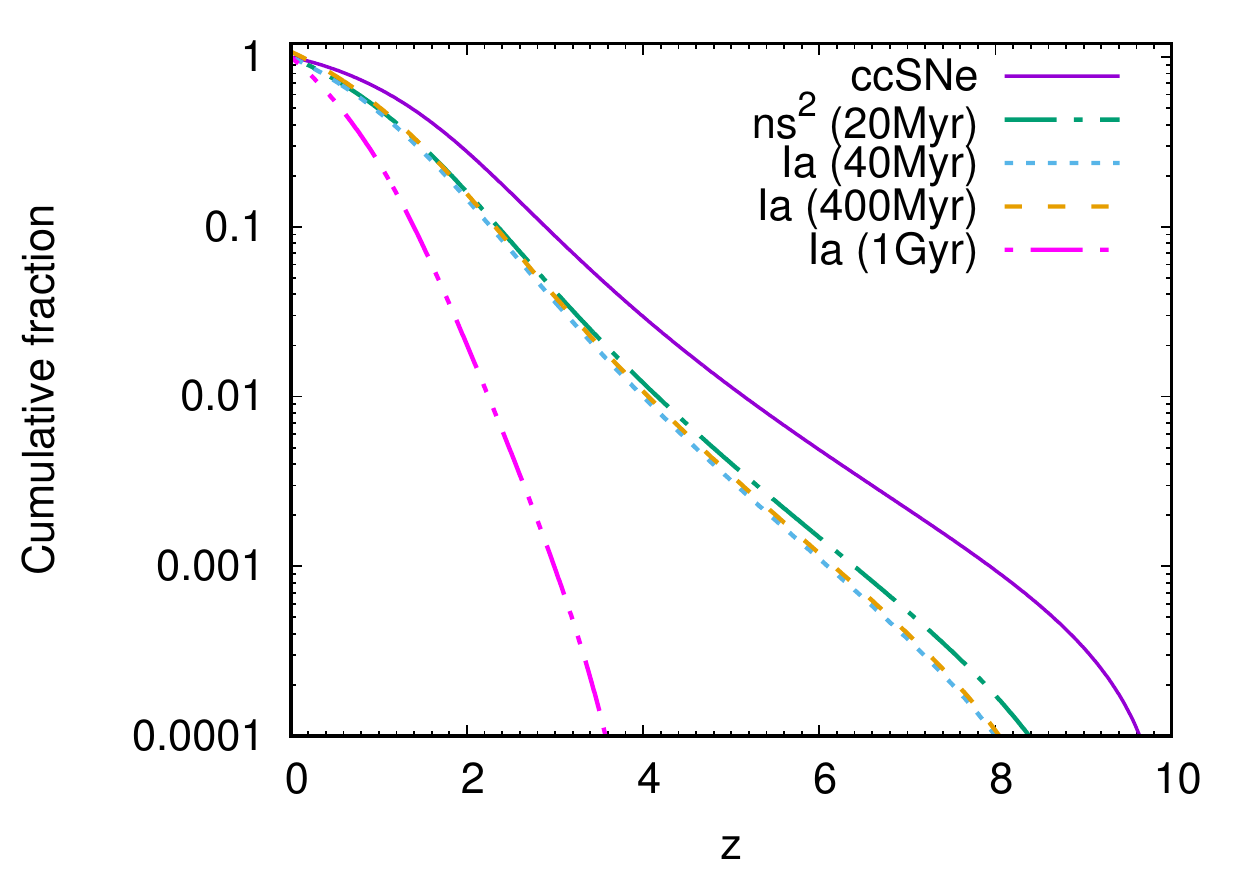}
\caption{The volumetric event rate of cc-SNe, SNe Ia, and neutron star mergers ({\it left})
and the cumulative fraction ({\it right}). The rate of cc-SNe is assumed to follow the cosmic star
formation history \citep{Madau2014}. SNe Ia and mergers are assumed to follow
it with  delay times distributed as $\propto t^{-1}$. The minimal delay time of each
model is shown in parentheses.
}
\label{fig:rate}
\end{figure*}

\begin{figure*}
\includegraphics[scale=0.6]{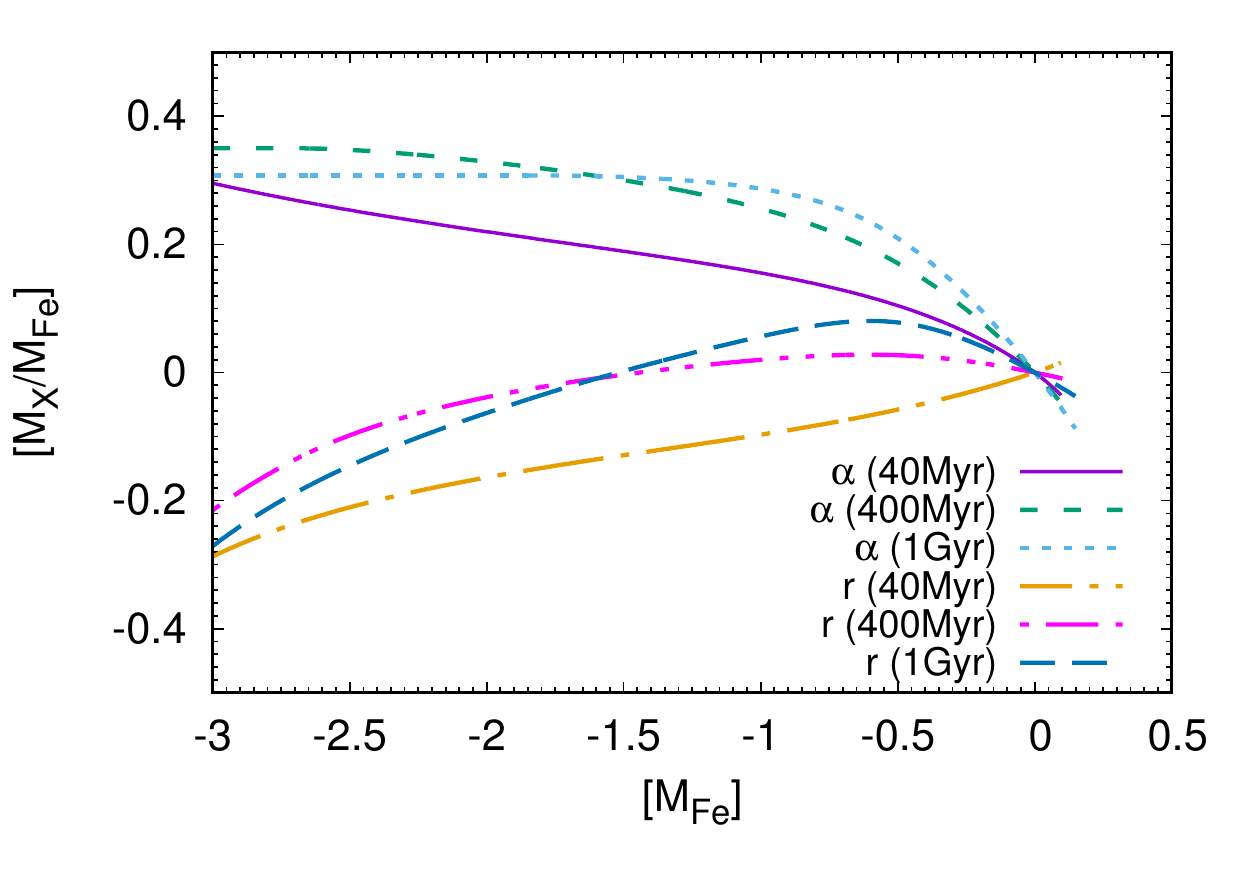}
\includegraphics[scale=0.6]{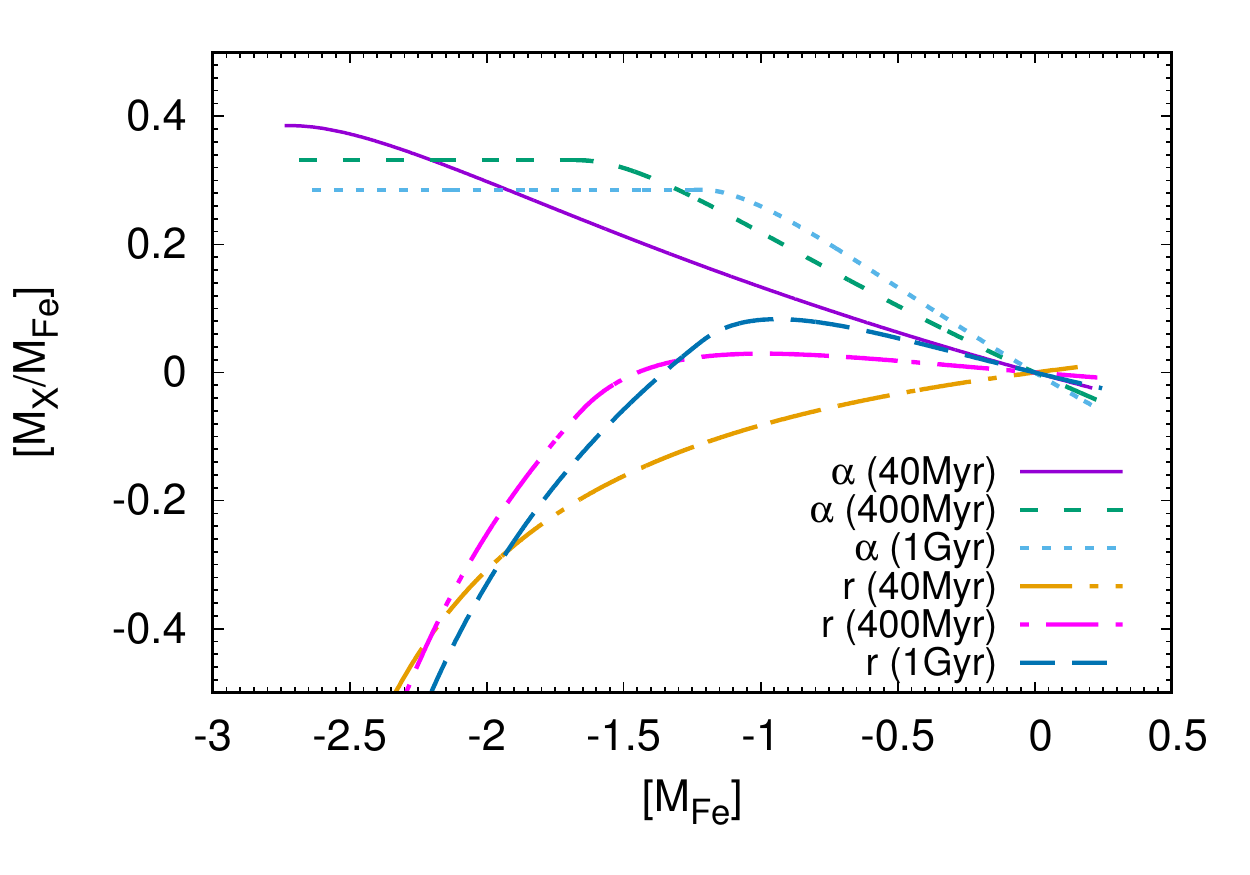}
\caption{The logarithmic ratio of the produced mass of $\alpha$ and $r$-process elements
as a function of the produced iron mass. The values in parentheses
show the minimal delay time of SNe Ia in each model.
The star formation history 
is assumed to follow the cosmic star formation history~({\it left}) or be constant over time~({\it right}).
}
\label{fig:ratio}
\end{figure*}

The event rate at a given time
is obtained by integrating $D_i(t)$ with the star formation rate. Figure \ref{fig:rate} shows the rate and cumulative fraction of cc-SNe following the cosmic
star formation rate \citep{Madau2014}, type Ia SNe with  minimal delay times  
of $40,\,400,\,10^3$\,Myr \citep{Totani2008,Maoz2014}, and neutron star mergers with $b_{\rm ns}=1$ and a minimal delay time of $20$\,Myr.   
Given these rates and the produced mass of each element per event, we calculate, following  \cite{Maoz2017},  the ratio of the total amount of elements produced by each phenomenon. 
Figure \ref{fig:ratio} shows [$M_{r}/M_{\rm Fe}$] and [$M_{\alpha}/M_{\rm Fe}$] as a 
function of [$M_{\rm Fe}/M_{\rm H}$], where we assume $\alpha$-elements are produced
by cc-SNe, iron is produced by cc-SNe and SNe Ia, and $r$-process elements are produced by neutron star mergers. These quantities are normalized such that  [$M_{X}/M_{\rm Fe}$] is zero 
at [$M_{\rm Fe}/M_{\rm H}$] $=0$. The different curves correspond to the different
minimal delay times of SNe Ia. 

[$M_{\alpha}/M_{\rm Fe}$] decreases  with [$M_{\rm Fe}/M_{\rm H}$] 
for [$M_{\rm Fe}/M_{\rm H}$]$>-1$. This  is consistent with metal poor star measurements as
shown in Fig. \ref{fig:abundance}. However, the decline rates of [$M_{r}/M_{\rm Fe}$] are always slower than those of $\alpha$-elements.  The production rate of $\alpha$-elements drops when 
the star formation rate drops. On the contrary, iron and $r$-process elements are still produced by
 SNe Ia and mergers that have long delay times $\gtrsim 1$\,Gyr. This feature can be also seen in the cumulative fraction of events  (the left panel 
of Fig. \ref{fig:rate}). In the neutron merger scenario, this slow decline of $r$-process elements compared to
 $\alpha$-elements is  inconsistent with the abundances of stars in the solar neighborhood.

\subsection{A simple one zone model}
We now turn to discuss the Galactic chemical evolution with a simple one zone model.
The total mass of $\alpha$-elements in the ISM is given  by 
\begin{eqnarray}
\frac{dM_{\alpha}}{dt} = m_{\alpha} C_{\rm cc} \Psi(t) - M_{\alpha} f(t) ,\label{alpha}
\end{eqnarray}
where $\Psi(t)$ is the star formation rate of the Milky Way (mass per unit time),
$m_{\alpha}$ is the mass of $\alpha$-elements produced per
cc-SN, $C_{\rm cc}$ is the cc-SN rate per unit mass, and
$f(t)$ is the mass-loss rate of the element in the ISM due to  
star formation and the galactic outflow.  
It is often assumed that the outflow rate is 
proportional to the star formation rate with a delay of the lifetimes of massive stars.
Since we consider time scales much longer than the lifetimes of massive stars
the outflow rate is considered to be proportional to the star formation rate:
\begin{eqnarray}
f(t) & = & ({\rm absorption\,rate\,by\, star\,formation\,+\,outflow\,rate})/M_{\rm ISM}\\
& = &(1+o) \frac{\Psi(t)}{M_{\rm ISM}},
\end{eqnarray}
where $o(t)$ is the outflow rate normalized by the star formation rate, which is likely smaller than unity,
 $M_{\rm ISM}$ is the
 total mass of the ISM. Here, we assume $M_{\rm ISM}\approx 10^{10}M_{\odot}$ at $z=0$.
 For the cosmic star formation history case, to estimate $M_{\rm ISM}$ as a function of time,
 we use the Kennicutt-Schmit relation  
  $\Psi \propto M_{\rm ISM}^{1.4}$ for a given $\Psi(t)$.
Using these relations, Eq. (\ref{alpha}) has simple solutions in the following two limits:
\begin{eqnarray} \label{f}
M_{\alpha}(t) \approx
 \left\{                                                                                                                                                                             
\begin{array}{ll} 
m_{\alpha}C_{\rm cc}M_*(t) ~~~~~~~~\,(M_* \ll M_{\rm ISM}),\\ 
\frac{m_{\alpha}C_{\rm cc}}{(1+o)}M_{\rm ISM}(t)~~~~~(M_* \gg M_{\rm ISM}),
\end{array}                                                                                                                                                                        
\right.
\end{eqnarray}
where $M_{*}$ is the total stellar mass.
At early times, the mass of $\alpha$-elements in the ISM increases  linearly with
the total stellar mass. Then, once 
the total stellar mass is larger than $M_{\rm ISM}$ the mass of $\alpha$-elements  approaches to a constant value.

For  neutron star mergers and for type Ia SNe, the production history
involves the delay time (relative to the star formation rate) distribution. Thus, 
the total mass of these elements  evolves differently from that of the 
$\alpha$-elements:
\begin{eqnarray}
\frac{dM_{\rm Fe}}{dt} & = &m_{\rm Fe,\,cc}C_{cc}\Psi(t) + m_{\rm Fe,\,Ia} \int_{0}^{t} dt' D_{\rm Ia}(t-t')\Psi(t') - M_{\rm Fe}(t) f(t),\\
\frac{dM_{r}}{dt} & = & m_{r} \int_{0}^{t} dt' D_{\rm ns}(t-t')\Psi(t') - M_{r}(t) f(t).
\end{eqnarray}
At late times $(M_* \gg M_{\rm ISM})$, the mass 
of $r$-process elements is given by
\begin{eqnarray}
M_{r}(t)\approx \frac{m_{r}}{(1+o)}\frac{\int_0^t D_{\rm ns}(t')\cdot \Psi(t')  dt'}{\Psi(t)}M_{\rm ISM}(t).
\end{eqnarray}
This increases logarithmically for $b_i=1$ if the star formation rate is constant. When we convert 
europium abundances to the total $r$-process abundances, we use the solar $r$-process abundance pattern.
As noted in \S \ref{sec:tot}, this procedure depends on the minimal atomic mass number
of $r$-process elements synthesized in mergers, $A_{\rm min}$. 
In the following, we use $A_{\rm min}=69$.

\subsection{A Comparison with Observations}
\begin{figure*}
\includegraphics[scale=0.6]{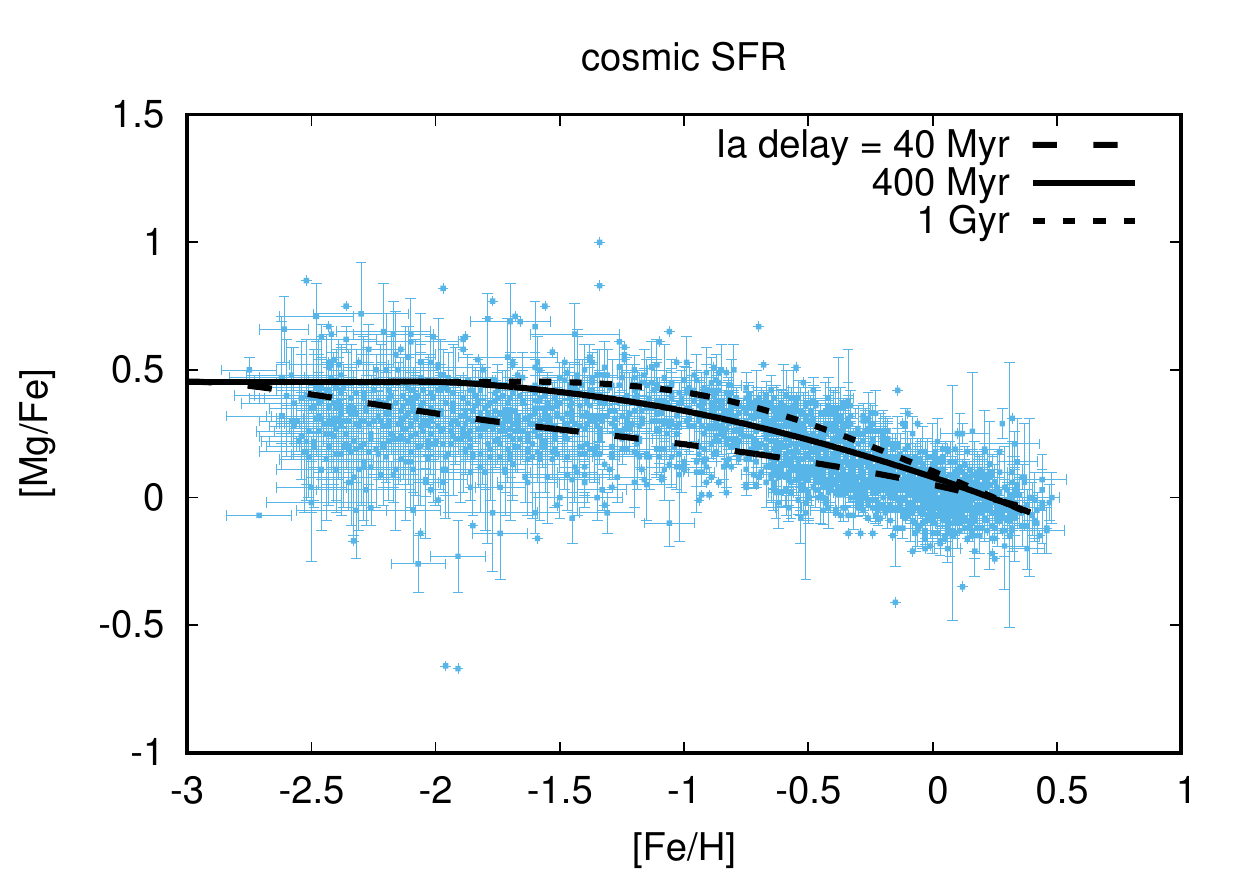}
\includegraphics[scale=0.6]{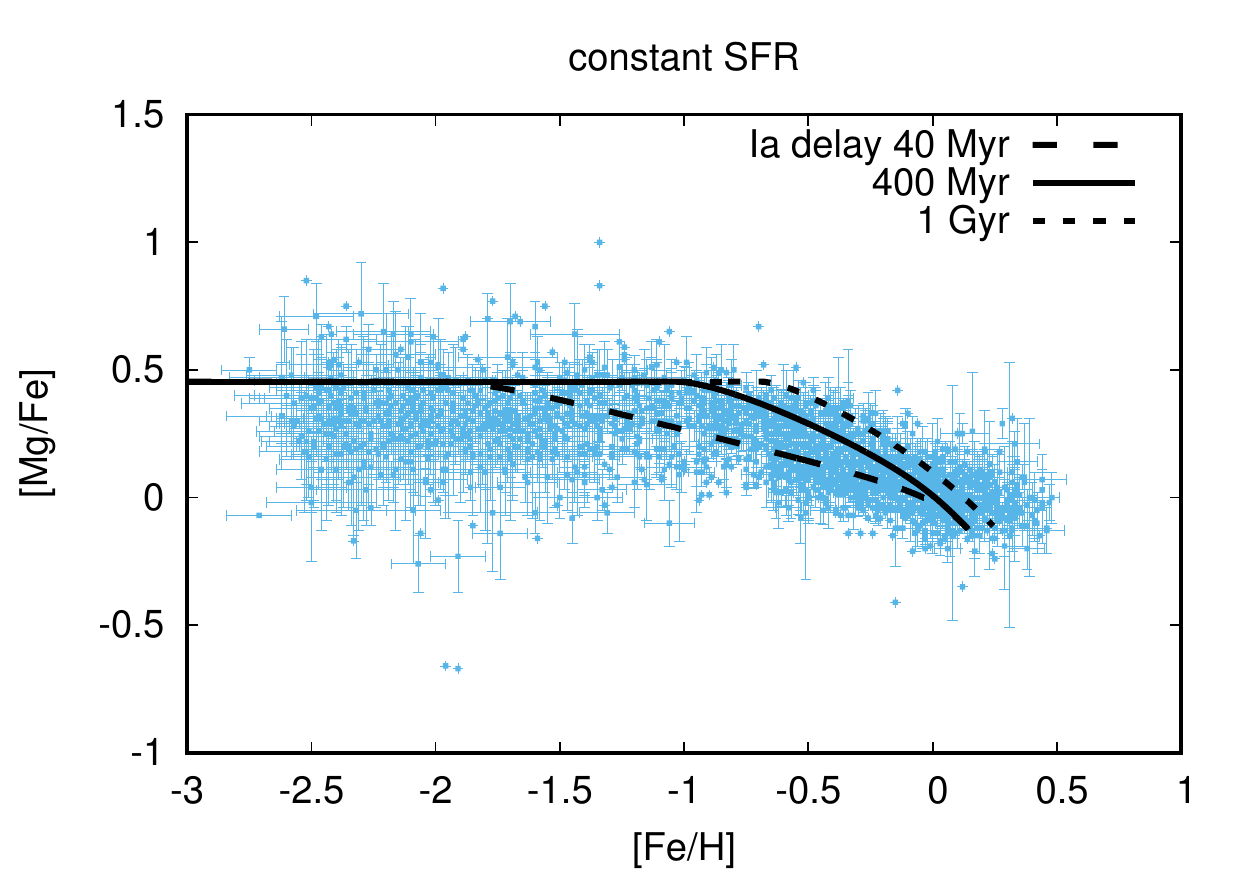}\\
\includegraphics[scale=0.6]{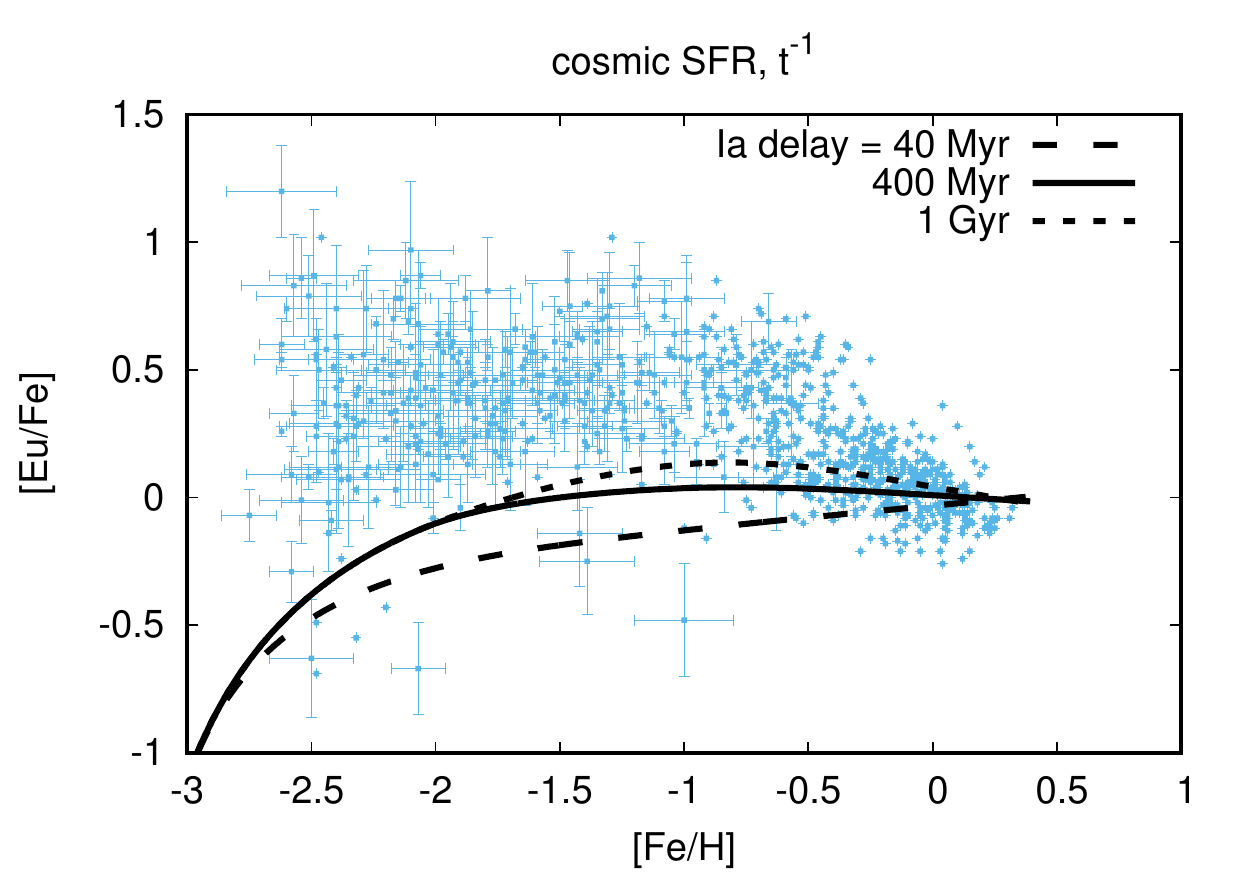}
\includegraphics[scale=0.6]{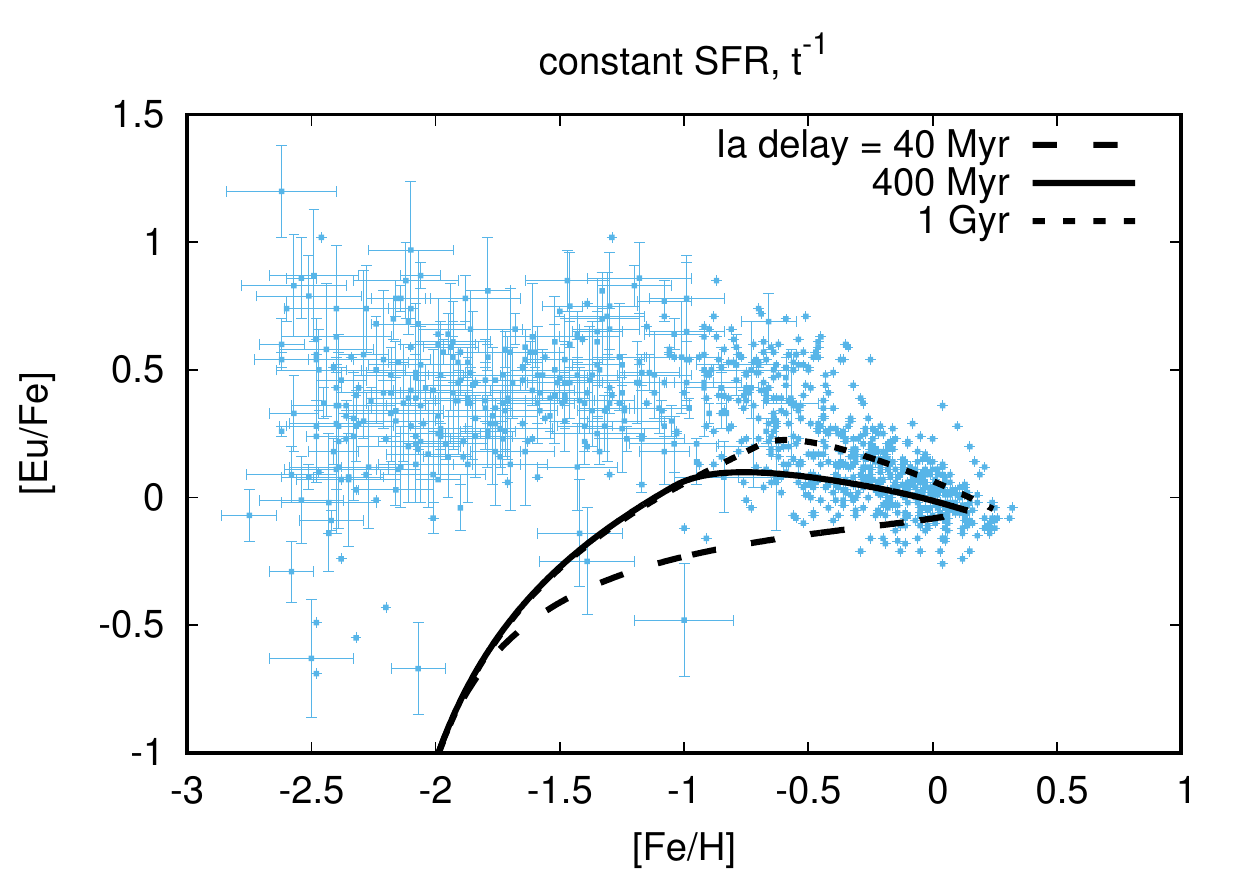}\\
\includegraphics[scale=0.6]{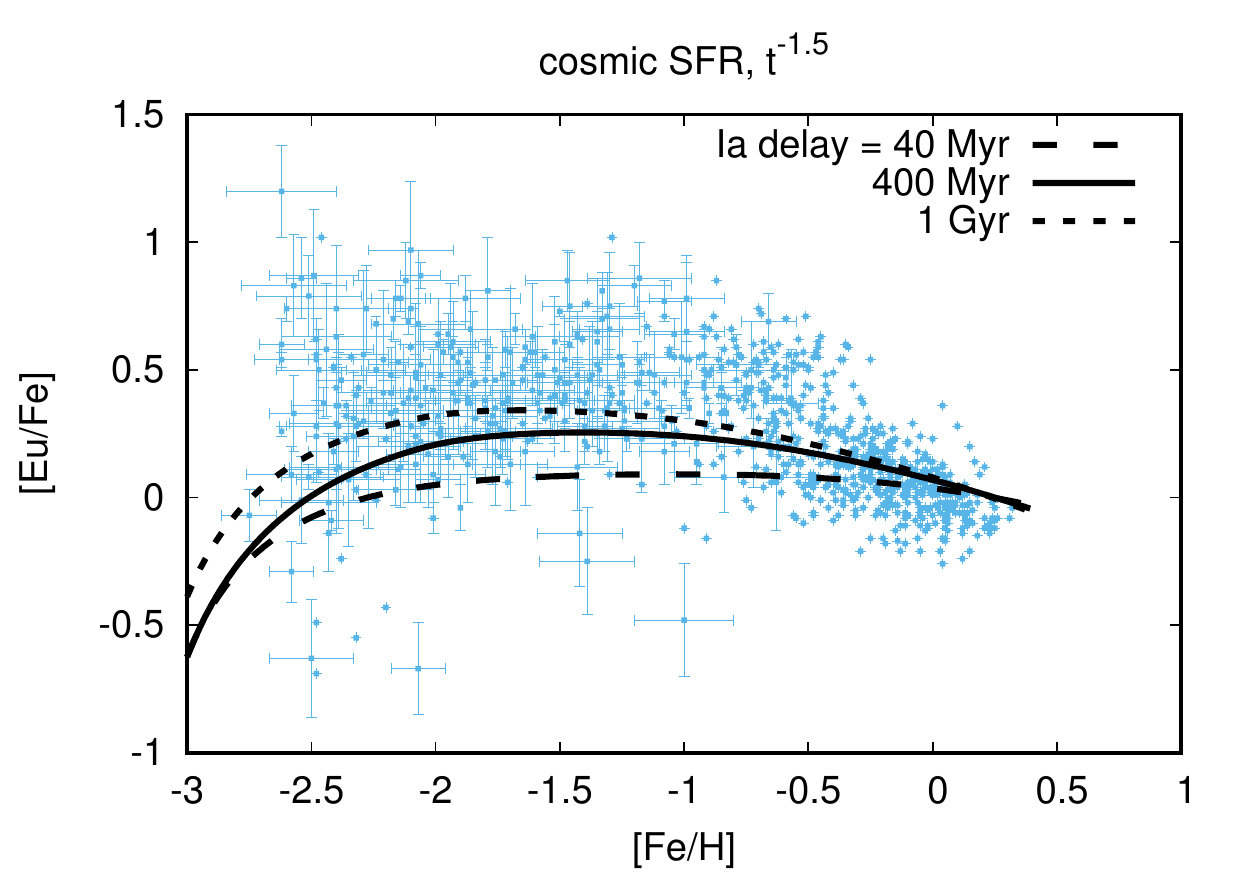}
\includegraphics[scale=0.6]{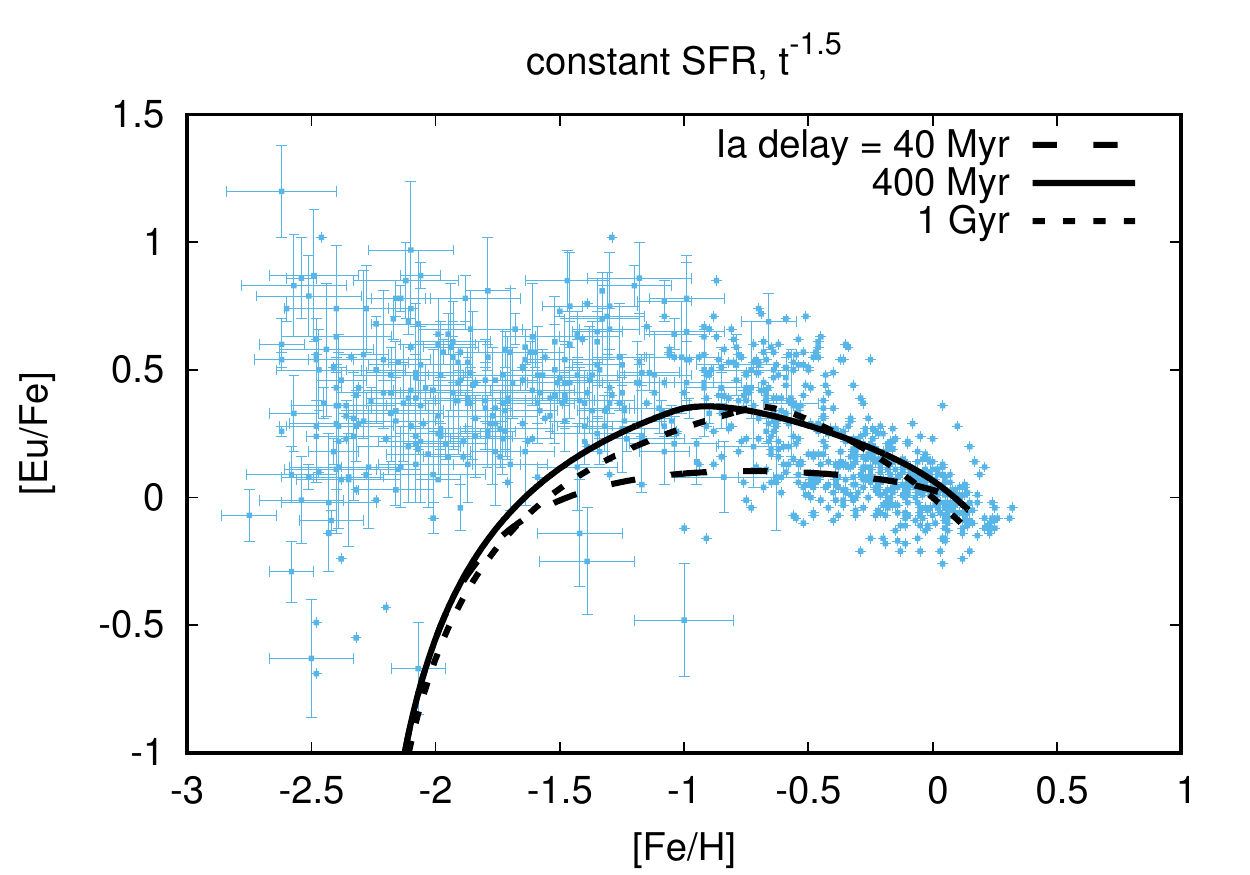}
\caption{
Magnesium ({\it top}) and europium ({\it middle} and {\it bottom}) abundance  evolutions of the Milky Way. 
Here, the minimum delay time of SNe Ia
is set to be $40$ (dashed), $400$ (solid), and $1000$~Myr (short dashed curve).   
The minimum delay time of neutron star mergers is chosen to be $20$~Myr and 
their delay time distribution is set to be $\propto t^{-1}$ ({\it middle}) and $\propto t^{-1.5}$ ({\it bottom}).
The star formation history 
is assumed to follow the cosmic star formation history~({\it left})
and to be a constant with time~({\it right}). 
 Also shown are the 
observed data of the magnesium and europium abundances of metal poor stars  
as  blue points from SAGA database \citep{suda2008}.
}
\label{fig:abundance}
\end{figure*}

Figure~\ref{fig:abundance} shows that the observed [Mg/Fe] and [Eu/Fe] for 
stars from SAGA database  (\citealt{suda2008} and references therein).
The mean value of the observed [Mg/Fe] is roughly a  constant
 up to [Fe/H] $\approx -1$.  Then it gradually decreases towards  its  value today,  $\approx 1/3$ of this constant. 
This is usually interpreted as due to the copious production of iron (without the production of  significant amounts of $\alpha$-elements) by  SNe Ia that  begin to take place at around  
[Fe/H] $\approx-1$ (e.g. \citealt{Nomoto2013} and references therein, but see also  \citealt{Maoz2017} for an alternative interpretation).

The resulting  evolutions of [Mg/Fe] and [Eu/Fe] are shown
in Fig.~\ref{fig:abundance}. Here, we examine two cases (i) a 
cosmic star formation history~(left panel) 
and  (ii) a constant star formation history ~(right panel).  
The motivation for considering   these star formation rate histories  is that interestingly the Galactic star formation rate history is not so clear. 
The cosmic star formation rate, mimics well the other possibility of an early ($\le 1$Gyr) star formation peak and one suggested approximation  is a constant star formation. 
The input parameters of each case are listed in 
Table~\ref{tab:model}.  Here, we fix the minimal delay time of neutron 
star mergers to be $20$ Myr and it extends to longer delay times as
$\propto t^{-b_{\rm ns}}$, where $b_{\rm ns}$ is set to be $1$ or $1.5$. 
 While $20$~Myr might look short, this minimal  delay time is consistent with the observed redshift distribution and peak flux 
of sGRBs \citep{WP15}. Those authors  find that the distribution of time delays $\propto t^{-0.8^{+0.25}_{-0.24}}$ for $t> 20 $~Myr is consistent with 
the observations. 
We show that the [Mg/Fe] and  [Eu/Fe] evolutions
with different minimal delay times of $40$, $400$, and $1000$~Myr for SNe Ia (see, e.g., \citealt{Totani2008,Maoz2014} for the SNe Ia delay time distribution).

 \begin{table}[h]
\caption[]{The input parameters for the one zone model.   A value in a parenthesis is for the constant SFR case.}
 {\begin{tabular}{lc} \hline \hline
 $m_{\rm Mg}$ & $0.12M_{\odot}/{\rm event}$  \\
 $m_{\rm Fe,\,cc}$ & $0.074M_{\odot}/{\rm event}$  \\
 $m_{\rm Fe,\,Ia}$ & $0.7M_{\odot}/{\rm event}$\\
 $m_{r}$ & $0.05\, (0.06)M_{\odot}/{\rm event}$  \\
  $C_{\rm cc}$ & $8\cdot 10^{-3}M_{\odot}^{-1}$\\
 $C_{\rm Ia}$ & $1.6\cdot10^{-3} M_{\odot}^{-1}$  \\
  $C_{\rm ns}$ & $7\cdot 10^{-6}M_{\odot}^{-1}$ \\
  $X_{\rm H}$ & $0.75$ \\
  $o$ & $0.25$ \\
\hline \hline \\
\label{tab:model}
\end{tabular}}

\end{table}
 
The expected decline rate of [Eu/Fe] is shallower than [Mg/Fe] because of the delay of mergers (Fig. \ref{fig:abundance}) relative to star formation.
[Eu/Fe] even increases with [Fe/H] if the minimal time delay of neutron 
star mergers is comparable to that of SNe Ia (see also \citealt{Cote16}).
For the constant star formation case, the decline rates for both [Mg/Fe] and [Eu/Fe]
are steeper than those for the cosmic one, but the decline of [Eu/Fe] 
is still slower than the observed one.
The bottom panels of Fig. \ref{fig:abundance} show
the evolution for a steeper delay time distribution,  $b_{\rm ns}=1.5$,  
for neutron star mergers. As a larger population of mergers follow star 
formation, the decline rate of [Eu/Fe] is more similar to [Mg/Fe],
suggesting that the declining rate of [Eu/Fe] of stars in the solar neighborhood 
can be explained with a somewhat steep delay time distribution 
and with that a minimal delay time of mergers are much shorter than that 
of SNe Ia. However, such a steep delay time distribution is inconsistent 
with the distribution, $\propto t^{-0.8\pm0.25}$, derived from the redshift distribution of sGRBs \citep{WP15}. {Furthermore,  if the merger rate following star formation 
is much higher than that with longer time delays,   it results in an overproduction of $r$-process elements in the Galaxy, given
the produced $r$-process mass of $\approx 0.05M_{\odot}$ and rate estimated from GW170817 (the time delay of GW170817 is  $\sim 1$--$10$ Gyr).
}

The neutron star merger rate used in this section corresponds to 
a Galactic rate of $\approx 10\,{\rm Myr^{-1}}$. This is slightly smaller than
the one inferred from GW170817. Note, however, that for the question of $r$-process nucleosynthesis the normalization of the 
merger rate this is degenerate
with the mass produced per event as discussed in \S \ref{sec:tot}. 
Furthermore, because of the natal kicks involved in the formation of
neutron star binaries,
a faction of neutron star mergers occur outside the Galactic disk
and they do not contribute to the Galactic chemical evolution.

\section{Summary and Discussions}
We have discussed that various measurements and theoretical estimates related to the  origin of  $r$-process elements including: the estimated rates of mergers and the estimates of yields of $r$-process material in each event,  the total abundance of heavy $r$-process elements in the Milky Way,  the abundance of radioactive elements in the   early solar system and the current deposition rate, the abundance of heavy $r$-process material in dwarf galaxies, and the recent observations of the kilonova/macronova associated with GW170817. We have shown that all these are compatible with the model in which  compact binary mergers are the dominant source of  heavy $r$-process elements in the Universe. 

Our main findings are:  
(i)  The rate and mass of $r$-process elements produced per event
inferred from  the europium abundances of Galactic stars, the geological
 abundances of radioactive elements, and the $r$-process enrichment  of dwarf 
 galaxies are consistent with each other. The inferred mass per event 
 is 
 $3\cdot 10^{-2}\lesssim m_r \lesssim 3\cdot 10^{-1}M_{\odot}$ for $A_{\rm min}=69$ and
$ 7\cdot 10^{-3}\lesssim m_r \lesssim 7\cdot 10^{-2}M_{\odot}$ for $A_{\rm min}=90$.
(ii)  The rate and mass of $r$-process elements produced per event
inferred from the optical-nIR signal (kilonova/macronova) of GW170817
are enough to explain the total amount of $r$-process elements in the Milky Way
and consistent with the other measurements. 
The rate and mass are also consistent with the rate of sGRBs and the ejecta mass
estimates  of the kilonova/macronova candidates associated with nearby sGRBs.
(iii)  The natal kicks of binary neutron stars and the time delay between the
formation and merger may reduce a population of mergers that contribute 
to the $r$-process enrichment, in particular, for small dwarf galaxies, of which 
the escape velocity is low and the star formation lasts for a relatively short time scale.

On the other hand, when we studied the Galactic chemical evolution, we encountered some problems 
with the simple one zone model. We compared the Galactic chemical evolution of iron, $\alpha$-elements,  and $r$-process elements (specifically europium) with a model  in which mergers follow the star formation rate with a delay time distribution, $\alpha$-elements are produced by cc-SNe and iron is produced predominantly by type Ia SNe but also by cc-SNe. 
The difficulty stems from the fact that the distribution of [Eu/Fe] for [Fe/H] $\gtrsim-1$ is quite similar to that of [$\alpha$/Fe].
Both fall by about a factor of 3 between [Fe/H] $\approx-1$ and the solar value. 
This indicates that {a significant fraction }  of both $\alpha$-elements and  $r$-process elements  is produced before SNe Ia begin to 
produce significant amounts of iron. The first is natural as we expect that $\alpha$-elements are produced by cc-SNe. 
However, this is somewhat problematic for  the neutron star merger scenario as $r$-process events are expected to be delayed relative to the cc-SNe. 
The expected theoretical distribution is $\propto t^{-1}$ (recall that this is also roughly the distribution implied by  sGRBs). In fact an intrinsic problem arises here since we have assumed that at late times both type Ia SNe and compact binary mergers follow the star formation rate convoluted with a $t^{-1}$ time delay distribution. This makes the late time decrease in [Eu/Fe] problematic. 
Unlike earlier works that suggest problems with early formation of $r$-process elements within the merger model we stress here a problem at late times (higher metallically problem). 

The model can fit the data if we stretch the parameters. Extending the time delay for the onset of type Ia SNe to 1Gyr, setting the minimal time delay of mergers at 20 Myr and employing a time delay distribution $\propto t^{-1.5}$.  In this case  a significant fraction of   mergers are pushed back to early times and the resulting [Eu/Fe] is consistent with observations. As mentioned earlier such a distribution is inconsistent with theoretical expectations and with sGRB observations, which show $t^{-0.80\pm.25}$ \citep{WP15}. { Furthermore, such a steep 
distribution combined with the rate, mass estimate, and time delay of
GW180817 seems to result in an overproduction of $r$-process elements in the Milky Way.}
 
This tension between the chemical evolution observations and our model is disturbing. However, before turning to conclude that mergers are inconsistent as sources of heavy $r$-process elements we should consider both the simplicity of our model and the complexity of the observed data.  

Concerning the observations we note that  there are several different samples of abundance measurements and it is not clear how to combine them. Furthermore, 
the abundances are compared to [Fe/H]. Especially at low metallicities  [Fe/H] may not vary at the same rate with time in different regions of the Galaxy. This latter fact is particularly important if the star formation history varies strongly in different regimes of the Galaxy. In fact, a larger sample of red giants in the APOGEE survey shows an $\alpha$-abundance distribution different from
that seen in the solar neighborhood ([$\alpha$/Fe] is lower for red giants at lower Galactic scale heights; \citealt{Hayden2015}). However, the $r$-process abundance distribution of those stars has not yet been determined.

As for the theoretical model, we have made some strong simplifying assumptions. First we have used one zone model with a single star formation history. It is clear that different regimes at the Galaxy (i.e. the bulge and the disk) formed at different times. Similarly we ignored the effect of merging dwarf galaxies { and the effect that mergers can occur in  low metallicity regions due to  natal kicks that launch neutron star binaries from one region in the Galaxy to another. } Depending on their star formation histories such events could have either enriched the Galaxy with $r$-process rich material or with very low metallically material \citep{hirai2015,shen2015ApJ,vandevoort2015MNRAS,ishimaru2015ApJ,Naiman2017,Hirai2017}.  The overall sign of this effect is not clear. Finally we have assumed that the type Ia rate is also dictated by gravitational radiation decay with a given minimal time delay followed by $t^{-1}$ distribution. While this is supported by observations \citep{Totani2008,Maoz2014}. Changing this time delay distribution to one dominated, for example, by the formation rate of white dwarfs (e.g. \citealt{Yungelson2000} but see also \citealt{Hachisu2008}) would change the [Fe/H] and yield a  different chemical evolutions that may resolve some of the problems found here.

{More neutron star mergers will be discovered in the upcoming observation runs of Advanced LIGO/Virgo. KAGRA and LIGO-India will also join in near future and adding these two detectors will improve the localization and the duty cycle. Identifying the host galaxies of mergers will reveal the event rate and delay time distribution of mergers. Furthermore, kilonovae/macronovae of different events may have different features, e.g., the ejecta mass and the fraction of lanthanides. 
In addition to binary neutron star mergers, GW signals from black hole-neutron star mergers are expected to be discovered. This will reveal the role of black hole-neutron star mergers for $r$-process material production in the Universe.    
Therefore, future GW and EM observations will give us better understanding on the r-process budget and on the variation in the abundance patterns of  $r$-rich extremely metal poor stars.
 }

We thank Chia-Yu Hu, Dan Maoz, Brian Metzger, Ehud Nakar, Yong-Zhong Qian, Masaru Shibata, Masaomi Tanaka, and Shinya Wanajo for useful discussions. 
K.H. was supported by the Flatiron Fellowship at the Simons Foundation and the Lyman Spitzer Jr. Fellowship awarded by Department of Astrophysical Sciences at Princeton University.
T.P. was partially supported by an advanced ERC grant TReX and by a grant from the Templeton foundation. We acknowledge kind hospitality at the Flatiron institute while some of this research was done.

\end{document}